\documentclass[fleqn]{aa}
\usepackage[dvips]{graphics}
\usepackage[psamsfonts]{euscript}
\usepackage[psamsfonts]{amssymb}
\usepackage{amsmath}

\newcommand{\dst}   {\displaystyle}
\newcommand{\rad}   {\mathrm{R}}
\newcommand{\Te}    {T_{\rm e}}

\newcommand{\Acal}  {{\EuScript A}}
\newcommand{\abf}   {\alpha_\mathrm{bf}(i,\nu)}
\newcommand{\aff}   {\alpha_\mathrm{ff}(\nu,\Te)}
\newcommand{\aijnu} {\alpha_{ij}(\nu)}
\newcommand{\aT}    {a_{\rm T}}
\newcommand{\aTa}   {a_{\rm T}^-}
\newcommand{\CCa}   {\EuScript{C}_0}
\newcommand{\CCb}   {\EuScript{C}_1}
\newcommand{\CCc}   {\EuScript{C}_2}
\newcommand{\Cij}   {C_{ij}}
\newcommand{\Cik}   {C_{i\kappa}}

\newcommand{\chiH}  {\chi_{\rm H}}
\newcommand{\cm}    {~\mathrm{cm}}
\newcommand{\compr} {\eta}
\newcommand{\divf}  {\nabla\!\cdot\mathbf{F}_\mathrm{R}}
\newcommand{\divfnu}{\nabla\!\cdot\mathbf{F}_\nu}
\newcommand{\divF}  {\frac{1}{\rho}\divf}
\newcommand{\dFz}   {\EuScript{D}}
\newcommand{\Ea}    {E_a}
\newcommand{\Eat}   {\widetilde{E}_a}
\newcommand{\Ee}    {E_\mathrm{e}}
\newcommand{\Eet}   {\widetilde{E}_\mathrm{e}}
\newcommand{\Eex}    {E_{\rm ex}}
\newcommand{\Eext}   {\widetilde{E}_{\rm ex}}
\newcommand{\Ei}     {E_{\rm I}}
\newcommand{\Eit}    {\widetilde{E}_{\rm I}}
\newcommand{\Erad}  {E_\rad}
\newcommand{\Erada} {E_\rad^-}
\newcommand{\Eradb} {E_\rad^+}

\newcommand{\Ett}   {\widetilde{E}_\mathrm{t}}
\newcommand{\Fcal}  {{\EuScript F}}
\newcommand{\Frad}  {F_\rad}
\newcommand{\Frada} {F_\rad^-}
\newcommand{\Fradb} {F_\rad^+}

\newcommand{\gcc}   {~{\rm gm}\:{\rm cm}^{-3}}
\newcommand{\Ha}    {H$\alpha$}
\newcommand{\intnu} {\int\limits_0^\infty}
\newcommand{\Jcal}  {{\EuScript J}}
\newcommand{\K}     {\mathrm{K}}
\newcommand{\kbf}   {\kappa_{\rm bf}(\nu)}
\newcommand{\kff}   {\kappa_{\rm ff}(\nu)}
\newcommand{\kl}    {\kappa_\ell(\nu)}
\newcommand{\kms}   {~{\rm km}~{\rm s}^{-1}}
\newcommand{\Lc}    {{L_\mathrm{c}}}
\newcommand{\Ll}    {{L_\ell}}
\newcommand{\Lya}   {Ly$\alpha$}
\newcommand{\Lyb}   {Ly$\beta$}
\newcommand{\me}    {m_{\mathrm e}}
\newcommand{\mH}    {m_{\mathrm H}}
\newcommand{\nel}   {n_{\rm e}}
\newcommand{\nela}  {n_{\rm e}^-}
\newcommand{\nelt}  {\tilde{n}_{\rm e}}
\newcommand{\nH}    {n_{\rm H}}
\newcommand{\nHa}   {n_{\rm H}^-}
\newcommand{\nni}   {n_i}
\newcommand{\nnilte}{\nni^*}
\newcommand{\nnit}  {\tilde{n}_i}
\newcommand{\nnj}   {n_j}
\newcommand{\nnjlte}{\nnj^*}
\newcommand{\nnjt}  {\tilde{n}_j}
\newcommand{\nk}    {n_\kappa}
\newcommand{\nklte} {\nk^*}
\newcommand{\Pa}    {P_a}
\newcommand{\Pe}    {P_{\rm e}}
\newcommand{\Pgas}  {P_{\rm g}}
\newcommand{\Pij}   {P_{ij}}
\newcommand{\Pik}   {P_{i\kappa}}
\newcommand{\Pji}   {P_{ji}}
\newcommand{\Pki}   {P_{\kappa i}}
\newcommand{\Prad}  {P_\rad}
\newcommand{\Prada} {P_\rad^-}
\newcommand{\Pradb} {P_\rad^+}
\newcommand{\Qea}   {Q_{{\rm e}a}}
\newcommand{\Qei}   {Q_{\rm ei}}
\newcommand{\Qelc}  {Q_{\rm elc}}
\newcommand{\Qinc}  {Q_{\rm inc}}
\newcommand{\RRa}   {{\mathfrak R}_a}
\newcommand{\RRe}   {{\mathfrak R}_\mathrm{e}}
\newcommand{\RRU}   {{\mathfrak R}_\mathrm{U}}
\newcommand{\Rij}   {R_{ij}}
\newcommand{\Rijd}  {\Rij^\dagger}
\newcommand{\Rik}   {R_{i\kappa}}
\newcommand{\Rikd}  {\Rik^\dagger}
\newcommand{\rhoa}  {\rho^-}
\newcommand{\rhob}  {\rho^+}
\newcommand{\sR}    {\sigma_\mathrm{R}(\nu)}
\newcommand{\sT}    {\sigma_\mathrm{T}}
\newcommand{\sumL}  {\sum\limits_{i=1}^L}
\newcommand{\sumLj} {\sum\limits_{\substack{j=1\\ j\ne i}}^L}
\newcommand{\Ta}    {T_a}
\newcommand{\Taa}   {T_a^-}
\newcommand{\Tab}   {T_a^+}
\newcommand{\Tea}   {T_{\rm e}^-}
\newcommand{\Teb}   {T_{\rm e}^+}
\newcommand{\Ua}    {U^-}
\newcommand{\Ub}    {U^+}
\newcommand{\xH}    {x_{\rm H}}
\newcommand{\Zik}   {Z_{i\kappa}}

\begin{document}

\thesaurus{02(02.19.1;02.08.1;02.18.7)}
\title{The structure of radiative shock waves}
\subtitle{II. The multilevel hydrogen atom}
\author{Yu.A.~Fadeyev\inst{1} \and D.~Gillet\inst{2}}
\institute{Institute for Astronomy of the Russian Academy of Sciences,
 Pyatnitskaya 48, 109017 Moscow, Russia \\
 email: fadeyev@inasan.rssi.ru
\and
 Observatoire de Haute-Provence - CNRS, F-04870
 Saint-Michel l'Observatoire, France \\
 email: gillet@obs-hp.fr }
\date{Received 3 September 1999 / Accepted 17 November 1999}
\titlerunning{The structure of radiative shock waves. II}
\maketitle

\begin{abstract}
Models of steady--state plane--parallel shock waves propagating
through the unperturbed hydrogen gas of temperature $T_1=6000\K$ and
density $\rho_1 = 10^{-10}\gcc$ are computed for upstream velocities
$15\kms\le U_1\le 70\kms$.
The properties of the ambient gas are typical for atmospheres of
pulsating stars.
The shock wave structure is considered in terms of the self--consistent
solution of the radiation transfer, fluid dynamics and rate equations
for $2\le L\le 4$ atomic bound levels with a continuum.
The radiative flux $\Frad$ emergent from the shock wave
was found to be independent of the lower limit $\nu_L$ of the frequency
range provided that $\nu_L < \nu_2$, where $\nu_2$ is the
Balmer continuum head frequency,
At the same time the decrease of $\nu_L$ is accompanied by
decrease of the Lyman continuum flux and leads to smaller heating and
weaker ionization of the hydrogen gas in the radiative precursor.
For all models the size of the radiative precursor is of $\sim 10^4\cm$
and corresponds to several mean free paths of photons at the frequency
of the Lyman continuum edge $\nu_1$.
The compression ratio at the discontinuous jump gradually increases with
increasing upstream velocity $U_1$, reaches the maximum of
$\rho^+/\rho^- = 3.62$ at $U_1\approx 55\kms$ and slowly decreases
for larger $U_1$ due to the strong rize of the preshock gas temperature.
The radiative flux from the shock wave was determined as
a function of the upstream velocity $U_1$ and its ratio to the total
energy flux in the shock wave $\CCc$ was found to range within
$0.18 < \Frad/\CCc < 0.92$ for $15\kms\le U_1\le 65\kms$.
Thus, at upstream velocities $U_1 > 60\kms$ the shock wave losses
more than 90\% of its total energy due to radiation.
For all shock wave models the role of collisional processes
in both bound--bound and bound--free atomic transitions was found to be
negligible in comparison with corresponding radiative processes.
\end{abstract}

\section{Introduction}

Attempts to obtain the self--consistent solution of the equations of fluid
dynamics and radiative transfer for shock waves propagating through
atmospheric gases were undertaken during several last decades.
The need to employ the self--consistent solution is of tremendous
importance in the case of the partially ionized preshock hydrogen gas
undergoing the substantial radiative heating and photoionization
due to the strong absorption of the Lyman continuum radiation
outgoing from the postshock region.
In order to correctly treat the coupling between the preshock and postshock
regions the two different methods known as the
``mean--photon approximation'' and the ``asymptotic layering method'' were
previously proposed (see, for recent review, Gillet 1999).
Unfortunately, in both these methods the radiative transfer is treated
as an initial value problem which is solved with use of the
shooting method.
The principal disadvantage of such an approach is that the true solution
of the transfer equation is unstable for optical depths $\tau > 1$
since it is affected by the exponentially growing errors.
This obstacle, however, can be overcome if the radiative transfer is solved
as a two--point boundary value problem.
In order to reach the self--consistency between the gas flow and the radiation
field Fadeyev \& Gillet (1998, hereinafter referred to as Paper~I)
employed the procedure of global iterations.
Each cycle of the global iteration procedure involves the solution of the
initial--value problem for the fluid dynamics and rate equations
which is followed by the solution of the two--point boundary value
problem for the radiation transfer equation.
As was shown in Paper~I, the global iterations converge
and provide with the stable self--consistent solution.

The goal of this study has been to further develop a method of global
iterations for calculating the structure of radiative shock waves.
In Paper~I the study was restricted to the model of the hydrogen
atom with two bound levels and a continuum.
The rate equations were solved for the first bound level and free
electrons, whereas the second atomic level was treated in LTE.
In our previous study we did not take into account also the rise of the
electron temperature due to the adiabatic compression at the
discontinuous jump.
Moreover, the convergence of global iterations has been reached
for shock models with insufficiently thick postshock region.
As a result, the models presented in Paper~I did not cover the full
recombination of the hydrogen and estimates of the total radiative
flux emergent from the shock wave remained uncertain.

In the present work the system of the ordinary differential equations
is extended by the terms taking into account the contribution of the
radiation field in the energy and momentum equations.
The rate equations are solved for all bound levels
of the adopted hydrogen atom model.
The substantial improvements of the computer program
allowed us to obtain the good convergence of global iterations for shock wave
models involving almost the entire recombination zone and,
thereby, to obtain the reliable estimates of the radiative flux emergent
from shock waves with upstream Mach numbers as large as $M_1\approx 7$.
As in Paper~I, the present study is confined to shock waves
propagating through the partially ionized hydrogen gas.

The present paper is organized as follows.
In Sect.~\ref{baseq} we describe details of our shock wave model,
give the fluid dynamics and rate equations written in the form
of the ordinary differential equations and
describe the solution of the transfer equation for the both continuum
and spectral line radiation.
In comparison with Paper~I the equations are completely rewriten
with the use of more convenient notations and should replace those of Paper~I
where numerous typographical
errors resulted from the missing of the \TeX\ macro definition file.
In Sect.~\ref{propsol} we discuss the properties of the solution,
demonstrate its convergence and degree of approximation.
General properties of the shock wave models are presented in
Sect.~\ref{descrmod}.
More detailed description of the models is given in
Sects.~\ref{radfld} -- \ref{postshreg} where we
discuss the properties of the radiation field and
the structure of the preshock and postshock regions, respectively.


\section{Basic equations}\label{baseq}

We consider steady--state, plane--parallel shock waves propagating through
the homogeneous medium composed solely of an atomic hydrogen gas.
The solution is determined as a function of three parameters:
the unperturbed gas temperature $T_1$, the unperturbed gas density $\rho_1$
and the velocity $U_1$ at which the shock wave moves with respect
to the ambient gas.
The model is represented by a flat finite slab comoving with the shock wave
and the origin is set at the infinitesimaly thin discontinuous jump
(adiabatic shock front) where
thermodynamic quantities undergo an abrupt change.

As was shown in Paper~I, the stable self--consistent solution
for the radiative shock wave structure can be obtained with the method
of global iterations.
In this method the rate equations for atomic level populations and
equations of fluid dynamics are treated as an initial--value problem
for the system of ordinary differential equations, whereas the radiation
transfer is solved in the framework of the two--point boundary value problem.
The interplay between the gas flow and the radiation field is taken into
account by the iteration procedure.
Such an approach allows us to substantially diminish the role of
the stiffness of the rate equations as well as to avoid large errors arising
due to the strong frequency dependence of the radiation field.

In order to apply the Feautrier technique (Feautrier 1964) for solution
of the transfer equation the slab is divided into $N-1$ cells
(see, e.g., Mihalas 1978).
The cell sizes increase in both directions from the discontinuous jump
locating at the $\Jcal$--th cell interface with space coordinate $X_\Jcal = 0$.
In cell interfaces with space coordinates $X_j$, where $1\le j\le N$,
we define the monochromatic optical depth $\tau_{\nu j}$ and the monochromatic
radiative flux $F_{\nu j}$ at frequency $\nu$.
Other variables are defined at the cell centers with space coordinates
\begin{equation}
X_{j-1/2} = \frac{1}{2}\left(X_{j-1} + X_j\right),\qquad (j=2,\ldots ,N).
\end{equation}
The outer boundaries of the preshock and postshock regions are
$X_1$ and $X_N$, respectively. For these boundary points we write
\begin{equation}
X_{1/2} = X_1 ,
\qquad
X_{N+1/2} = X_N
\end{equation}
since all variables must be also defined at both boundaries.

Integration of the ordinary differential equations begins at the
preshock outer boundary $X_1$ where the gas is assumed to be unperturbed
and the number densities of free electrons $\nel$ and bound atomic levels
$\nni$ are in statistical equilibrium with radiation field of the shock wave.
Preshock integration is done on the interval $[X_1, X_{\Jcal -1/2}]$,
where $X_{\Jcal -1/2}$ is the space coordinate of the cell center just
ahead the discontinuous jump.
Initial conditions for the postshock integration are obtained from the
Rankine--Hugoniot equations relating variables at the cell centers
$X_{\Jcal -1/2}$ and $X_{\Jcal +1/2}$.
The outer boundary of the postshock region $X_N$ is set behind
the recombination zone.
When integration of the ordinary differential equations on the interval
$[X_{\Jcal +1/2}, X_N]$ is completed,
we solve the transfer equation for the whole slab.
Below we discuss the steps of the global iteration procedure in more detail.

\subsection{Rankine--Hugoniot relations} 

The zonal quantities at the cell centers $X_{\Jcal -1/2}$ and $X_{\Jcal +1/2}$
are related by the radiation modified Rankine--Hugoniot equations (Marshak 1958)
\begin{equation}
\label{C0}
\rho U = \CCa \equiv \dot m ,
\end{equation}
\begin{equation}
\label{C1}
\dot m U + \Pgas + \Prad = \CCb ,
\end{equation}
\begin{equation}
\label{C2}
\dot m \left(h + \frac{1}{2}U^2\right) + \Frad + U\left(\Erad + \Prad\right) =
 \CCc ,
\end{equation}
where
$\Erad$, $\Frad$ and $\Prad$ are the radiation energy density,
radiation flux and radiation pressure, respectively,
\begin{equation}
\label{pgas}
\Pgas = \Pa + \Pe = \nH k\Ta + \nel k\Te
\end{equation}
is the gas pressure,
$\nH$ and $\Ta$ are the number density and the temperature of hydrogen atoms
both neutral and ionized, $\Te$ is the temperature of free electrons.
The specific enthalpy is
\begin{equation}
\label{enth}
h = \frac{5}{2}\frac{\nH}{\rho}k\Ta + \frac{5}{2}\frac{\nel}{\rho}k\Te +
 \frac{\Eex}{\rho} + \frac{\Ei}{\rho} ,
\end{equation}
where
\begin{equation}
\Eex = \chiH\sumL\left(1 - i^{-2}\right)\nni
\end{equation}
is the excitation energy,
$L$ is the number of bound atomic levels,
$\Ei = \chiH\nel$ is the ionization energy,
$\chiH$ is the ionization potential of the hydrogen atom.

Eqs.~(\ref{C0}) -- (\ref{C2}) are solved for the compression ratio
\begin{equation}
\label{eta}
\compr = \frac{\rhob}{\rhoa} = \frac{\Ua}{\Ub} ,
\end{equation}
where superscripts minus and plus denote for brevity
the quantities defined at cell centers $X_{\Jcal -1/2}$ and $X_{\Jcal +1/2}$,
respectively.

Substituting (\ref{pgas}) and (\ref{enth}) into (\ref{C1}) and (\ref{C2})
and then combining two resulting equations we obtain
\begin{equation}
\label{quadeq}
A\compr^2 - B\compr + C =0 ,
\end{equation}
where
\begin{equation}
\label{Aterm}
A = \left(\aTa\right)^2 +
    {\left(\Ua\right)^2\over 5} +
    \frac{2}{5}{\Frada - \Fradb\over\dot m} +
    \frac{2}{5}{\Erada + \Prada\over\rhoa} ,
\end{equation}
\begin{equation}
\label{Bterm}
B = \left(\aTa\right)^2 + \left(\Ua\right)^2 +
    {\Prada - \Pradb\over\rhoa} -
    \frac{2}{5}{\Eradb + \Pradb\over\rhoa} ,
\end{equation}
\begin{equation}
\label{Cterm}
C = \frac{4}{5}\left(\Ua\right)^2 ,
\end{equation}
and $\aT = \left(\Pgas/\rho\right)^{1/2}$ is the isothermal sound speed.

The postshock electron temperature $\Teb$ is determined in the
assumption of the adiabatic compression of the electron gas:
\begin{equation}
\Teb = \compr^{\gamma - 1}\Tea ,
\end{equation}
where $\gamma = \frac{5}{3}$ is the ratio of specific heats.
The postshock temperature of heavy particles is
\begin{eqnarray}
\label{Tab}
\Tab &=& \Taa + \frac{\nela}{\nHa}\left(\Tea - \Teb\right) +
\frac{1}{5}{\dot m\Ua\over\nHa k}
\left(1 - \frac{1}{\compr^2}\right) + \nonumber\\
&+&\frac{2}{5}{\Frada - \Fradb\over\nHa k U^-}
 + \frac{2}{5}{\Erada - \Prada - \Eradb - \Pradb\over\nHa k} .
\end{eqnarray}

The last two terms in (\ref{Aterm}), (\ref{Bterm}) and
(\ref{Tab}) take into account the contribution of the radiation field
which rapidly increases with increasing upstream velocity.
However, for models considered in the present study this contribution
was found to be rather small.
For example, at the upstream velocity of $U_1 = 70\kms$ omitting of these
terms leads to the increase of the compression ratio $\compr$ by
nearly one percent and to the increase of the postshock temperature $\Tab$
by less than one tenth of the percent.
Nevertheless, we retain these terms in Eqs.~(\ref{C1}) and (\ref{C2})
in order to use the momentum flux $\CCb$
and the total energy flux $\CCc$ for checking the accuracy of the model
throughout the slab.

\subsection{Fluid dynamics and rate equations} 

The solution vector of the system of ordinary differential equations is
\begin{equation}
\mathbf Y = \{\tilde{n}_1, \ldots , \tilde{n}_L, \nelt, \Eat, \Eet, U\} ,
\end{equation}
where $\Ea = \frac{3}{2}\nH k\Ta$ and $\Ee = \frac{3}{2}\nel k\Te$
are the translational energies of hydrogen atoms and free electrons
per unit volume, respectively.
Here and below the tilde denotes the quantities expressed per unit mass,
that is, $\nnit = \nni/\rho$, $\Eat = \Ea/\rho$ etc.

The assumption of the one--dimensional steady flow in planar geometry
implies that throughout the slab the continuity equation is
written as $\rho U = \dot m \equiv \CCa$.
Thus, the system of the fluid dynamics and rate equations is
\begin{eqnarray}
\label{dnidt}
\frac{d\nnit}{dt} &=& \sumLj \nnjt\Pji + \nelt\Pki - \nnit\sumLj\Pij - \nnit\Pik ,\\
\label{dnedt}
\frac{d\nelt}{dt} &=& \sumL \nnit\Pik - \nelt\sumL\Pki , \\
\label{dEadt}
\frac{d\Eat}{dt} &=& -\Pa\frac{dV}{dt} - \Qelc , \\
\label{dEedt}
\frac{d\Eet}{dt} &=& -\Pe\frac{dV}{dt} + \Qelc + \Qinc -
\frac{1}{\rho}\frac{d\Erad}{dt} , \\
\label{dUdt}
\frac{dU}{dt} &=& -\frac{1}{\dot m}\frac{d\Pgas}{dt} +
\frac{1}{\rho c}\int\limits_0^\infty\chi_\nu F_\nu\, d\nu ,
\end{eqnarray}
where $\Pij$ and $\Pji$ are the total (collisional plus radiative) rates
of bound--bound upward and downward transitions between the $i$--th and
$j$--th atomic levels, respectively,
$\Pik$ and $\Pki$ are the total rates of ionizations and recombinations for
the $i$--th atomic level,
$\Qelc$ and $\Qinc$ are the rates of energy gain by electron gas in
elastic and inelastic collisions, respectively,
$\chi_\nu$ is the monochromatic extinction coefficient,
$V=1/\rho$ is the specific volume,
$c$ is the velocity of light.

Eqs.~(\ref{dnidt}) -- (\ref{dUdt}) are written for time derivatives
in their left--hand--sides,
whereas for calculation of the shock wave spatial structure
it is more convenient to integrate the differential equations
with respect to the space coordinate $X$.
Furthermore, the time derivatives of the gas pressure and specific volume
should be expressed in terms of integrated variables.
Differentiating (\ref{pgas}) and (\ref{C1}) we have
\begin{equation}
\label{dPgdt}
\frac{d\Pgas}{dt} = \frac{2}{3}\rho
\left(\frac{d\Eat}{dt} + \frac{d\Eet}{dt}\right) -
\rho\Pgas\frac{dV}{dt} .
\end{equation}
and
\begin{equation}
\label{dVdt1}
\frac{dV}{dt} = -\frac{1}{\dot m^2}
\left(\frac{d\Pgas}{dt} + \frac{d\Prad}{dt}\right) .
\end{equation}
Substituting (\ref{dPgdt}) and
\begin{equation}
\label{dPrdt}
\frac{d\Prad}{dt} = -\frac{U}{c}\intnu\chi_\nu F_\nu\,d\nu .
\end{equation}
into (\ref{dVdt1}) we  write the time derivative of the specific volume as
\begin{equation}
\label{dVdt2}
\frac{dV}{dt} = -\Acal\left(\frac{d\Eat}{dt} + \frac{d\Eet}{dt}\right) + \Fcal ,
\end{equation}
where
\begin{equation}
\Acal = \left[\frac{3}{2}\dot m U\left(1 - \beta^2\right)\right]^{-1} ,
\end{equation}
\begin{equation}
\Fcal = {1\over\dot m\left(1 - \beta^2\right)\rho c}\intnu\chi_\nu F_\nu d\nu ,
\end{equation}
and $\beta = \aT/U$.

With (\ref{dPgdt}) and (\ref{dVdt2}) Eqs.~(\ref{dnidt}) -- (\ref{dUdt})
can be rewritten as
\begin{eqnarray}
\label{dnidX}
U\frac{d\nnit}{dX} &=&
\sumLj\nnjt\Pji + \nelt\Pki - \nnit\sumLj\Pij - \nnit\Pik , \\
\label{dnedX}
U\frac{d\nelt}{dX} &=& \sumL\nnit\Pik -\nelt\sumL\Pki , \\
\label{dEadX}
U\frac{d\Eat}{dX} &=& {\Acal\Pa\over 1 - \Acal\Pgas}\Qinc - \Qelc - U\RRa , \\
\label{dEedX}
U\frac{d\Eet}{dX} &=& {1 - \Acal\Pa\over 1 - \Acal\Pgas}\Qinc + \Qelc - U\RRe , \\
\label{dUdX}
\frac{dU}{dX} &=&-\Acal\dot m\frac{d\Eat}{dX} -\Acal\dot m\frac{d\Eet}{dX} + \RRU ,
\end{eqnarray}
where the terms
\begin{eqnarray}
\RRa &=&
\left(
{1 - \Acal\Pe\over 1 - \Acal\Pgas}\Pa + {\Acal\Pa\over 1 - \Acal\Pgas}\Pe
\right)
{\Fcal\over U} +
\nonumber\\
& & +
{\Acal\Pa\over 1 - \Acal\Pgas}{1\over\dot m}{d\Erad\over dt} ,
\end{eqnarray}
\begin{eqnarray}
\RRe &=&
\left(
{\Acal\Pe\over 1 - \Acal\Pgas}\Pa + {1 - \Acal\Pa\over 1 - \Acal\Pgas}\Pe
\right)
{\Fcal\over U} +
\nonumber\\
& & +
{1 - \Acal\Pa\over 1 - \Acal\Pgas}{1\over\dot m}{d\Erad\over dt} ,
\end{eqnarray}
and
\begin{equation}
\RRU = \rho\Fcal
\end{equation}
take into account the contribution of the radiation field and
can be omitted if $\Erad\ll \Ea + \Ee$ and $\Prad\ll\Pgas$.

The excitation rate per atom in the $i$--th initial state is
\begin{equation}
\label{Pij}
\Pij = \nel\Cij + \Rij
\end{equation}
and the de--excitation rate per atom in the $j$--th initial states is
\begin{equation}
\label{Pji}
\Pji = \frac{\nnilte}{\nnjlte}\left(\nel\Cij + \Rijd\right) ,
\end{equation}
where $\nnilte$ and $\nnjlte$ are the bound--level number densities
of hydrogen atoms in $i$--th and $j$--th state given by
the Saha--Boltzmann equation for the actual non--equilibrium number density
of free electrons $\nel$.

The collisional excitation rate is
\begin{equation}
\label{Cij}
\Cij = {h^2\over (2\pi\me)^{3/2} k^{1/2}}
{\sum\limits_{ll'}^{}\Gamma_{iljl'}\over g_j\Te^{1/2}}
\exp\left(\frac{\chiH}{i^2} - \frac{\chiH}{j^2}\right) ,
\end{equation}
where $g_j$ is the statistical weight of level $j$ and
$\Gamma_{iljl'}$ is the effective collision strength.
In the present study the effective collision strengths were evaluated from
analytic fits by Scholz \& Walters (1991) for $1s-2$ transitions,
Callaway \& Unnikrishnan (1993)  for $1s-3$ transitions and
Callaway (1994)  for $2-3$ transitions.
For other transitions the effective collision strengths were calculated
using the Chebyshev polynomial fits of the data by Aggarwal et al. (1991).

The photoexcitation rates are given by
\begin{equation}
\Rij = 4\pi\intnu{\aijnu\over h\nu} J_\nu d\nu ,
\end{equation}
where $\aijnu$ is the spectral line absorption cross--section in
transition $i\to j$ and $J_\nu$ is the monochromatic mean intensity.
The rate of downward (spontaneous plus induced) radiative transitions
$j\to i$ is
\begin{equation}
\frac{\nnilte}{\nnjlte}\Rijd = \frac{\nnilte}{\nnjlte}4\pi\intnu{\aijnu\over h\nu}
\left({2h\nu^3\over c^2} + J_\nu\right)
\exp\left(-{h\nu\over k\Te}\right) .
\end{equation}

The total rate of ionizations from the $i$--th level is
\begin{equation}
\Pik = \nel\Cik + \Rik
\end{equation}
and the total rate of recombinations to the $i$--th level is
\begin{equation}
\Pki = \frac{\nnilte}{\nklte}\left(\nel\Cik + \Rikd\right) ,
\end{equation}
where $\nklte$ is the number density of hydrogen ions evaluated from the
Saha--Boltzmann equation for the actual non--equilibrium number density
of free electrons $\nel$.
The collisional ionization rate is
\begin{equation}
\Cik = \nel\pi a_0^2 \left({8k\over\me\pi}\right)^{1/2} \Te^{1/2}
\exp\left(-{\chi_i\over k\Te}\right)\Gamma_i(\Te) ,
\end{equation}
where $\chi_i$ is the potential of ionization from the $i$--th level,
the $\Gamma(\Te)$ is a slowly--varying function of $\Te$
evaluated with fitting formulae given by Mihalas (1967).
The rates of phoionizations and photorecombinations are
\begin{equation}
\Rik = 4\pi\intnu {\abf\over h\nu} J_\nu d\nu
\end{equation}
and
\begin{equation}
\frac{\nnilte}{\nklte}\Rikd = \frac{\nnilte}{\nklte}4\pi\intnu
{\abf\over h\nu}\left({2h\nu^3\over c^2} + J_\nu\right)e^{-h\nu/k\Te}d\nu ,
\end{equation}
where $\abf$ is the bound--free absorption cross section at frequency $\nu$
by hydrogen atoms in the $i$--th state.
As is shown below, the role of collisional processes in the both
bound--bound and bound--free transitions throughout the radiative shock wave
is negligible since their rates are smaller by several orders of magnitude
in comparison with rates of radiative transitions.

The rate of energy gain by electrons in inelastic collisions is (Murty 1971)
\begin{equation}
\label{qinc}
\Qinc = - \frac{d\Eit}{dt} - \frac{d\Eext}{dt} - \divF ,
\end{equation}
where
\begin{equation}
\label{divf}
\divf = 4\pi\intnu\left(\eta_\nu - \kappa_\nu J_\nu\right) d\nu
\end{equation}
is the divergence of radiative flux,
$\eta_\nu$ and $\kappa_\nu$ are the emission and absorption coefficients,
respectively.

Behind the discontinuous jump the temperature of heavy particles $\Ta$
exceeds the electron temperature $\Te$ and free electrons acquire
the energy in elastic  collisions with heavy particles.
The rate of energy gain in elastic collisions is a sum of
the rates corresponding to collisions with hydrogen ions
and neutral hydrogen atoms:
\begin{equation}
\Qelc = \Qei + \Qea .
\end{equation}
The rate of energy gain in elastic collisions with hydrogen
ions is (Spitzer \& H\"arm 1953)
\begin{equation}
\Qei = \frac{2}{3}\frac{\nel}{\rho}k{\Ta - \Te\over t_{\rm eq}}
\,,
\end{equation}
where
\begin{equation}
t_{\rm eq} = {252\Te^{3/2}\over\nel\ln\Lambda}
\end{equation}
is the equipartition time and
\begin{equation}
\Lambda = 9.43 + 1.15\log\left(\Te^3/\nel\right) .
\end{equation}
The rate of energy gain by electrons in elastic collisions with neutral
hydrogen atoms is
\begin{equation}
\Qea = \frac{\nel}{\nH}\frac{\me}{\mH}n_1{\Ta - \Te\over\Te}
\langle\sigma_{{\rm e}a}v^3\rangle ,
\end{equation}
where the elastic scattering cross section is (Narita 1973)
\begin{eqnarray}
&&\langle\sigma_{{\rm e}a}v^3\rangle = \int\limits_0^\infty
\sigma_{{\rm e}a}v^3 f(v)\, dv =
\nonumber \\
&&=4\pi a_0^2\left(\frac{8}{\pi}\right)
\left(\frac{k\Te}{\me}\right)^{3/2}
\left[
4 + {24\over\left(1 + 2\cdot 10^{-5}\Te\right)^3}
\right] .
\end{eqnarray}

\subsection{Radiation transfer equation} 

The quasi--static radiation transfer equation for the planar geometry is
\begin{equation}
\label{rteq1}
\mu\frac{dI_{\mu\nu}}{d\tau_\nu} = I_{\mu\nu} - S_\nu ,
\end{equation}
where $I_{\mu\nu}$ is the specific intensity of radiation
at the directional cosine $\mu = \cos(\theta)$ at frequency $\nu$,
$S_\nu = \eta_\nu/\kappa_\nu$ is the monochromatic source function and
$\tau_\nu$ is the monochromatic optical depth defined as
$d\tau_\nu = \chi_\nu dX$.
The absoprtion and extinction coefficient are
\begin{equation}
\kappa_\nu = \kbf + \kff + \kl
\end{equation}
and
\begin{equation}
\chi_\nu = \kappa_\nu + n_1\sR + \nel\sT
\,,
\end{equation}
where $\kbf$ is the bound--free absorption coefficient,
$\kff$ is the free--free absorption coefficient,
$\sR$ is the cross section of the Rayleigh scattering by hydrogen atoms in
the ground state, $\sT$ is the Thomson scattering cross section,
$\kl$ is the spectral line absorption coefficient.
Expressions for $\kbf$ and $\kff$ can be found, for example, in the
book by Mihalas (1978).
The Rayleigh scattering cross section $\sR$ was evaluated using the fitting
formulae given by Kurucz (1970).

The continuum emission coefficient is
\begin{equation}
\eta_\nu = {2h\nu^3\over c^2}e^{-h\nu/ k\Te}
\left[\sumL\nnilte\abf + \nel^2\aff\right] ,
\end{equation}
where
\begin{equation}
\aff = {4e^6\over 3ch}
\left({2\pi\over 3k\me^3}\right)^{1/2}
{g_\mathrm{ff}(\nu,\Te)\over\nu^3\sqrt{\Te}}
\end{equation}
is the free--free absorption cross section and $g_\mathrm{ff}(\nu,\Te)$
is the free--free Gaunt factor.

The spectral line absorption and emission coefficients are
computed in the assumption of the complete redistribution, that is,
\begin{equation}
\kl = {\nni B_{ij} h\nu_{ij}\over 4\pi}
\left(1 - {\nnj g_i\over\nni g_j}\right)\phi_\nu
\end{equation}
and
\begin{equation}
\eta_\ell(\nu) = {h\nu_{ij}\over 4\pi}A_{ji}\nnj\phi_\nu ,
\end{equation}
where $g_i$ and $g_j$ are statistical weights,
$A_{ji}$ and $B_{ij}$ are Einstein coefficients,
$\phi_\nu$ is the spectral line absorption profile.
Thus, we take into account the frequency dependence of the source function
$S_\ell(\nu)=\eta_\ell(\nu)/\kl$ within the spectral line profile.

The radiative transfer problem is subject to the boundary conditions
implying that the incident radiation on both faces of the slab is due to the
thermal equilibrium radiation from the ambient unperturbed medium:
\begin{equation}
\label{bndc}
I_{\mu\nu 1/2}^+ =I_{\mu\nu N+1/2}^- = B_\nu(T_1) ,
\end{equation}
where $B_\nu(T_1)$ is the Planck function,
$T_1$ is the temperature of the unperturbed gas and superscripts minus and plus
correspond to $\mu < 0$ and $\mu > 0$, respectively.
The transfer equation (\ref{rteq1}) with boundary conditions (\ref{bndc})
is solved for the symmetric intensity average
\begin{equation}
u_{\mu\nu j-1/2} = \frac{1}{2}\left(I_{\mu\nu j-1/2}^+ + I_{\mu\nu j-1/2}^-\right)
\end{equation}
defined in the angle range $0\le\mu\le 1$ (Feautrier 1964).
The main difficulty arising in the solution of the transfer equation for
the shock wave is due to the very small optical depth intervals
$\Delta\tau_\nu$ in the vicinity of $X_\Jcal = 0$ at frequencies lower than
the ionization threshold of the level $i=3$. In the present study
this difficulty was overcome with use of the improved Feautrier solution
proposed by Rybicki \& Hummer (1991) which provides the better
numerical conditioning and prevents the lost of machine accuracy
in the matrix elimination scheme.

In each cell center the intensity average $u_{\mu\nu j-1/2}$ is determined
for sets of frequency and angle points.
In order to compute the frequency integrals the continuum frequency range
is divided into intervals with boundaries at the threshold
ionization frequencies $\nu_i$.
The upper boundary of the continuum frequency range is $\nu_0$.
Continuum intervals are represented by $n_\mathrm{c}$ nodes of
the Gauss--Legendre quadratures and spectral lines are represented by
$n_\ell$ nodes.
The radiation energy density $\Erad$, radiative flux $\Frad$ and radiation
pressure $\Prad$ are determined by summation of frequency integrals
over continuum intervals and over spectral lines:
\begin{equation}
\label{Erad}
\Erad =
\sum\limits_{i=1}^\Lc\frac{4\pi}{c}\int\limits_{\nu_i}^{\nu_{i-1}} J_\nu d\nu +
\sum\limits_{i=1}^\Ll\frac{4\pi}{c}\int\limits_{\nu_i-\Delta\nu_i}^{\nu_i+\Delta\nu_i} J_\nu d\nu ,
\end{equation}
\begin{equation}
\label{Frad}
\Frad =
\sum\limits_{i=1}^\Lc 4\pi\int\limits_{\nu_i}^{\nu_{i-1}} H_\nu d\nu +
\sum\limits_{i=1}^\Ll 4\pi\int\limits_{\nu_i-\Delta\nu_i}^{\nu_i+\Delta\nu_i} H_\nu d\nu ,
\end{equation}
\begin{equation}
\label{Prad}
\Prad =
\sum\limits_{i=1}^\Lc\frac{4\pi}{c}\int\limits_{\nu_i}^{\nu_{i-1}} K_\nu d\nu +
\sum\limits_{i=1}^\Ll\frac{4\pi}{c}\int\limits_{\nu_i-\Delta\nu_i}^{\nu_i+\Delta\nu_i} K_\nu d\nu ,
\end{equation}
where $\Lc$ is the number of frequency intervals, $\Ll$ is the number of
spectral lines, $\Delta\nu_i$ is the half--width of the $i$--th spectral line
frequency interval.
The moments of the radiation field in Eddington's notation are
\begin{equation}
\label{Jnu}
J_\nu = \int\limits_0^1 u_{\mu\nu}d\mu ,
\end{equation}
\begin{equation}
\label{Hnu}
H_\nu = \int\limits_0^1 v_{\mu\nu}\mu d\mu ,
\end{equation}
\begin{equation}
\label{Knu}
K_\nu = \int\limits_0^1 u_{\mu\nu}\mu^2 d\mu ,
\end{equation}
where
\begin{equation}
v_{\mu\nu} =-\mu{du_{\mu\nu}\over d\tau_\nu}
           = \frac{1}{2}\left(I_{\mu\nu}^+ - I_{\mu\nu}^-\right)
\end{equation}
is the antisymmetric flux--like average.


\section{Properties of the solution}\label{propsol}

The thickness of the adiabatic compression region where thermodynamic variables
undergo an abrupt change is only a few particle mean free paths.
Therefore, in comparison with the whole shock wave the adiabatic compression
region can be treated to a high degree of approximation as an infinitesimaly
thin discontinuous jump.
The assumption of the steady--state flow allows us to represent the shock
wave by the flat finite slab comoving with the discontinuous jump, so that
the integration of the fluid dynamics and rate equations follows the
element of gas which passes through the slab.
The axis of the space coordinate $X$ is chosen in such a way that the velocity
of the gas element is $U = dX/dt > 0$ and
the radiative flux emergent from the preshock and
postshock outer boundaries is $F_1 < 0$ and $F_N > 0$, respectively.

For all models the space coordinate of the preshock outer
boundary is $X_1 = -10^5$~cm, that is the size of the preshock region
exceeds by an order of magnitude the size of the radiative precursor.
The gas element crosses the preshock outer boundary at temperature $T_1$,
density $\rho_1$ and velocity $U_1$.
The radiative field at the preshock outer boundary is non--equilibrium,
so that the number densities of hydrogen atoms in $i$--th state $\nni$
and the number density of free electrons $\nel$ should be evaluated from
the solution of the equations of statistical equilibrium rather than
in LTE approximation.
To this end we solve Eqs.~(\ref{dnidt}) and (\ref{dnedt})
with time derivatives equated to zero using the Newton--Raphson
iteration procedure (see, e.g., Mihalas 1978).
It should be noted, however, that the number densities
obtained from the solution of the equations of statistical equilibrium
do not differ substantially from
their counterparts given by the Saha--Boltzmann equation.
For example, the hydrogen ionization degree at the preshock outer boundary
is in the range $8.9\cdot 10^{-3} < \xH < 2.5\cdot 10^{-2}$ for upstream gas
flow velocities $15\kms\le U_1 \le 70\kms$, whereas the LTE hydrogen
ionization degree is $\xH = 8.4\cdot 10^{-3}$.
Departure coefficients
\begin{equation}
b_i = {\nni/\nnilte\over\nk/\nklte}
\end{equation}
of the first three levels change from
$\mathbf{b} = \left\{0.9, 1.1, 1.1\right\}$ to
$\mathbf{b} = \left\{0.1, 0.3, 0.9\right\}$
for the same range of $U_1$.

Of great importance is the choice of the space coordinate $X_N$
of the postshock outer boundary.
In particular, the slab should involve the layers
of the full hydrogen recombination since otherwize the total radiative flux
emergent from the shock is underestimated.
Test calculations show that for the postshock outer boundary
located behind the maximum of the hydrogen ionization
the total radiative flux asymptotically approaches its limiting
value with increasing $X_N$.
At the same time, the increase of $X_N$ is accompanied by the growth of the
oscillation amplitude of iterated variables in the outermost layers
of the postshock region,
so that for too large $X_N$ the global iterations ultimately diverge.
This occurs mostly due to the limited accuracy of integration
of the ordinary differential Eqs.~(\ref{dnidX}) -- (\ref{dUdX}).
In the present study the space coordinate $X_N$ was determined for each
shock wave model from trial computations and ranged from
$X_N = 5\cdot 10^5$~cm for $U_1 = 70\kms$ to
$X_N = 6\cdot 10^7$~cm for $U_1 = 15\kms$.

In order to solve the radiation transfer equation (\ref{rteq1})
with the Feautrier technique
the slab is divided into $4000\le N\le 6000$ cells with
1500 cells in the preshock region.
The cells are smallest at $X_\Jcal = 0$ and increase in both
directions from the discontinuous jump according to the geometrical progression.
For all models the cell size at the discontinuous jump is
$\Delta X_{\Jcal -1/2} = \Delta X_{\Jcal +1/2} = 0.1$~cm.
The continuum frequency range is divided into $\Lc = L$ frequency intervals
$[\nu_i,\nu_{i-1}]$, where $\nu_i$ is the threshold frequency for ionization
from the $i$--th state, $1\le i\le L$ and $\nu_0 = 10^{16}$~Hz
is the the upper boundary of the frequency range.
The frequency integrals (\ref{Erad}) -- (\ref{Prad}) were computed
using the Gauss--Legendre quadratures with $n_\mathrm{c} = 12$
nodes for the continuum
intervals and with $n_\ell = 11$ nodes for spectral lines.
For the angular integration of the Eddington moments (\ref{Jnu}) -- (\ref{Knu})
we make use of the Gauss--Legendre quadrature with $n_\mu=12$ nodes.

In order to consider the dependence of the solution on the number of
bound atomic levels $L$ and on the width of the frequency range
$[\nu_L,\nu_0]$ we calculated three sequences of the models
with $L = \Lc = 2$, 3 and 4.
The unperturbed temperature and density of the ambient hydrogen gas
are $T_1 = 6000\K$ and $\rho_1 = 10^{-10}\gcc$, respectively.
These conditions are typical for atmospheres of Cepheids and RR Lyr
type variables.
The convergence of global
iterations is very sensitive to the initial approximation,
so that the first model of each sequence with $U_1 = 15\kms$
(the upstream adiabatic Mach number is $M_1 \approx 1.6$)
was computed using the initial LTE approximation with
postshock temperatures $\Ta$ and $\Te$ exponentially decaying
with increasing distance from the discontinuous jump
over the e--folding distance of $10^6$~cm.
Each converged model was used as an initial approximation for the following
model of higher amplitude. In order to avoid strong initial
oscillations of iterated quantites the models were computed with
amplitude increment $\Delta U_1 = 1\kms$.
The convergence of global iterations depends also on the size of cells
approximating the postshock region, so that is why the shock models
were represented by several thousand cells.

\begin{figure}
\resizebox{\hsize}{!}{\includegraphics{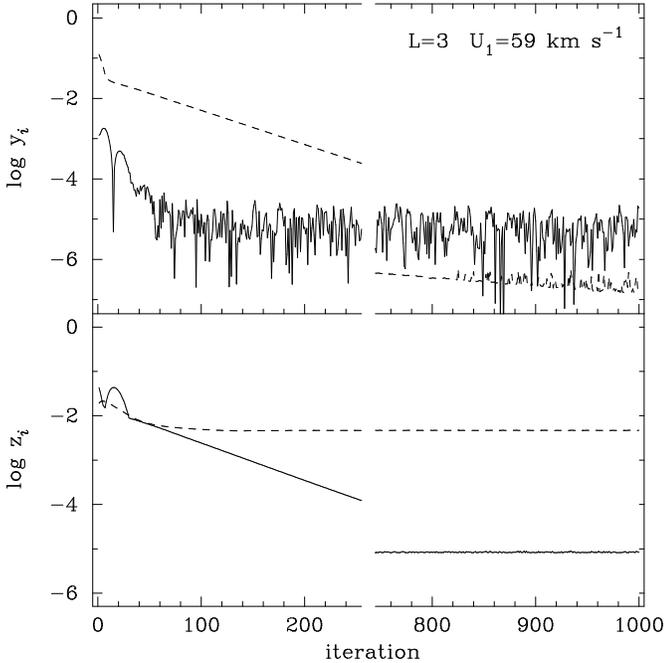}}
\caption{Convergence plots as a function of the number of global iterations
for the model with $U_1 = 59\kms$.
Upper panel: the maximum corrections of the radiative flux emergent
from the slab (solid line) and the electron temperature (dashed line).
Lower panel: the maximum relative change of $\CCb$ (solid line)
and $\CCc$ (dashed line) throughout the slab.}
\label{u59gi}
\end{figure}

Hydrodynamic variables that are most sensitive to the accuracy of the solution
are the electron temperature $\Te$ and the temperature of heavy particles
$\Ta$ in the vicinity of the postshock outer
boundary as well as the radiation flux $\Frad$ emergent from the slab.
Therefore, the convergence of the solution on the $(k+1)$--th iteration
was controlled by the maximum correction of the electron
temperature within the slab
\begin{equation}
y_1 = \max\limits_{1\le j\le N+1}
\left|1 - T_{{\rm e}j-1/2}^{(k)}/T_{{\rm e}j-1/2}^{(k+1)}\right|
\end{equation}
and by the correction of the radiative flux $F_1$ emergent from the
outer boundary of the preshock region
\begin{equation}
y_2 = \left|1 - F_1^{(k)}/F_1^{(k+1)}\right|
\,.
\end{equation}
If global iterations converge, both $y_1$ and $y_2$ exhibit the exponential
decay with increasing number of global iterations $k$.
The typical behaviour of $y_1$ and $y_2$  is displayed in Fig.~\ref{u59gi}.
After a few hundred iterations these quantities cease to decay and
oscillate around the constant average values
due to the limited accuracy of computations.
In the present study the stability of the converged solution was checked
for some models within several thousand iterations.

For controlling the degree of approximation we evaluated also
in all cell centers of the model
the momentum flux $\CCb$ and the total energy flux $\CCc$ defined by
Eqs.~(\ref{C1}) and (\ref{C2}), respectively.
By definition, both these quantities should remain constant throughout the slab
if the solution is exact.
On the lower panel of Fig.~\ref{u59gi} are shown the maximum deviations
\begin{equation}
z_i = \max\limits_{1\le j\le N+1}\left|1 - \EuScript{C}_{ij-1/2}/\EuScript{C}_{i1}\right|,
\quad (i=1,2),
\end{equation}
where $\EuScript{C}_{i1}$ are defined at  the preshock outer boundary $j=1$
and $\EuScript{C}_{ij-1/2}$ are defined at the cell centers
$1\le j\le N+1$, respectively.
As is seen from the lower panel of Fig.~\ref{u59gi}, the total energy flux
$\CCc$ is constant throughout the slab
with errors of less than a few tenths of a percent.


\section{General description of models}\label{descrmod}

\begin{figure}
\resizebox{\hsize}{!}{\includegraphics{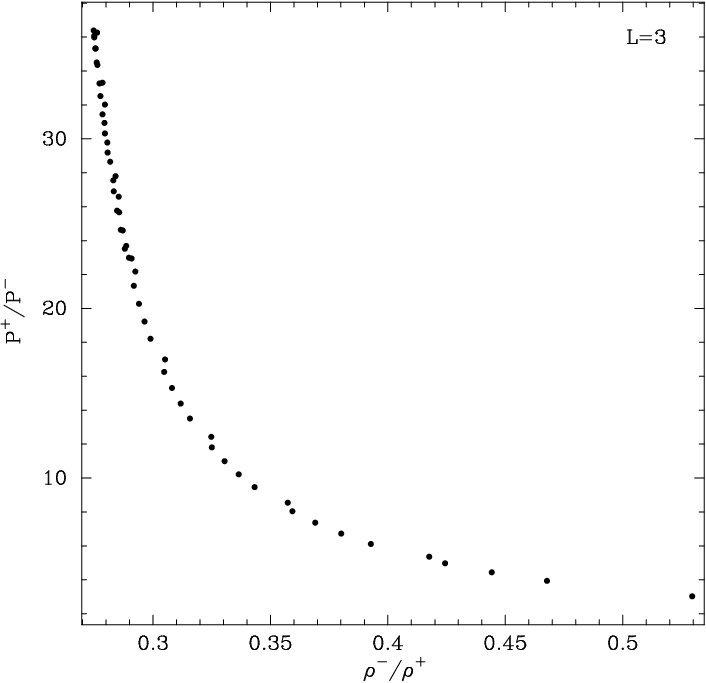}}
\caption{The pressure ratio $P^+/P^-$ versus the inverse
compression $\compr^{-1} = \rhoa/\rhob$ for models with $L=3$
and $15\kms\le U_1\le 70\kms$.}
\label{hugonc}
\end{figure}

Shock wave models computed with the same
upstream gas temperature $T_1$ and gas density $\rho_1$
can be instructively represented by the
Hugoniot curve (see, e.g., Zeldovich \& Raizer 1966) relating
the total pressure $P$ and the specific volume $V=1/\rho$ just ahead
and just behind the discontinuous jump.
In our case such a representation can be only approximate because both the
gas density and the total pressure ahead the discontinuous jump
grow with increasing upstream velocity due to radiative heating
in the preshock region. In particular, for upstream velocities
$15\kms\le U_1\le 70\kms$ the gas density increases by as much as
two percent, whereas the total pressure increases by nearly
a factor of three.
It should be noted that the growth of the preshock gas density
and the gas pressure becomes
substantial only at upstream velocities $U_1 > 50\kms$.
In comparison with the amplidute of the jump these changes, however,
are enough small.
The Hugoniot diagram for the sequence of models with $L=3$ is shown in
Fig.~\ref{hugonc}. The state of the gas at the cell center $X_{\Jcal -1/2}$
is given for all models by $(1,1)$ and is not shown on the plot.
The state of the gas just behind the discontinuous jump at the
cell center $X_{\Jcal +1/2}$ is represented for each model
by the filled circle.

\begin{figure}
\resizebox{\hsize}{!}{\includegraphics{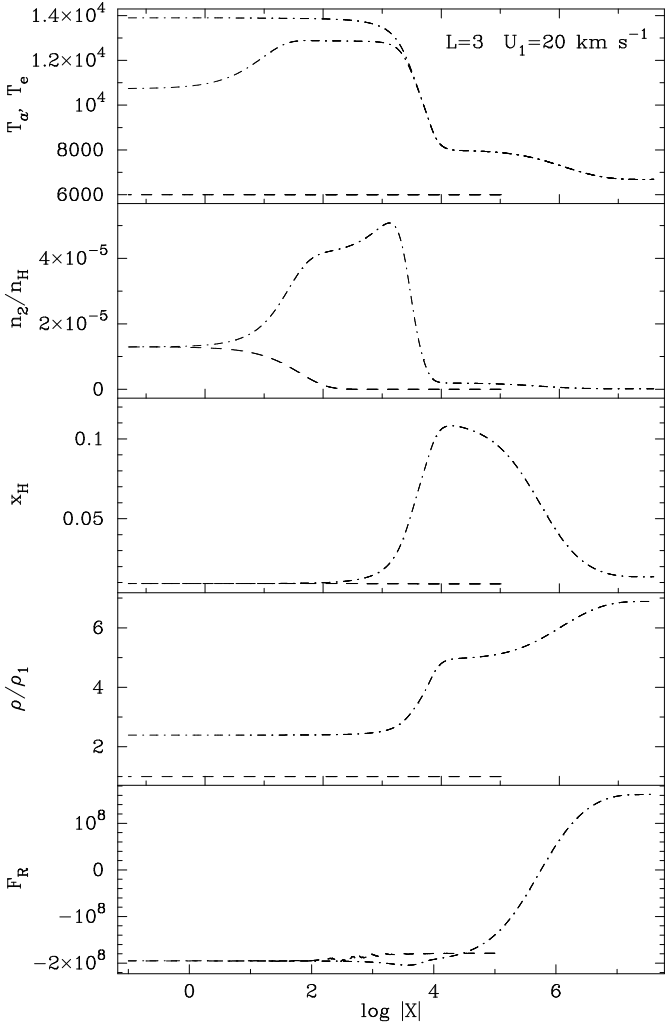}}
\caption{The electron temperature $\Te$, the temperature of heavy particles
$\Ta$, the ratio of the population number density of the second bound level
to the number density of hydrogen atoms $n_2/\nH$,
the degree of hydrogen ionization $\xH$,
the compression ratio $\rho/\rho_1$,
and the total radiative flux $\Frad$ versus the distance from the
discontinuous jump $X$ for the model with $L=3$, $U_1 = 20\kms$.
In dashed and dot--dashed lines are given the plots for
the preshock and postshock regions, respectively.}
\label{u20}
\end{figure}

\begin{figure}
\resizebox{\hsize}{!}{\includegraphics{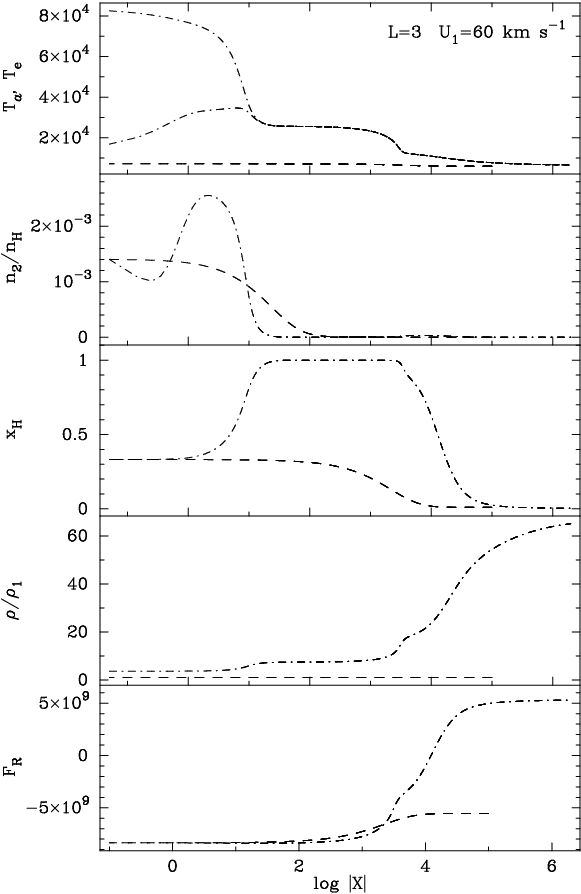}}
\caption{Same as Fig.~\ref{u20} for model with $L=3$, $U_1 = 60\kms$.}
\label{u60}
\end{figure}

General properties of converged models computed for $L = 3$
and upstream velocities $U_1 = 20$, 40, $60\kms$ are displayed in
Figs.~\ref{u20}\,--\,\ref{u60}, where the electron temperature $\Te$,
the temperature of heavy particles $\Ta$,
the ratio of the number density of hydrogen atoms in the second state to the
number density of hydrogen atoms $n_2/\nH$,
the hydrogen ionization degree $\xH$,
the compression ratio $\rho/\rho_1$, and the total radiative flux $\Frad$
are shown as a function of the distance from the discontinuous jump $X$.
For the sake of convenience the independent variable $X$ is in logarithmic
scale and each plotted variable is represented by two branches
corresponding to the preshock and postshock regions, respectively.
Just behind the discontinuous jump the temperature of neutral hydrogen atoms
and hydrogen ions $\Ta$ exceeds the temperature of the electron gas $\Te$.
That is why in the upper panel of Figs.~\ref{u20}\,--\,\ref{u60}
the postshock temperature plots are represented by two converging curves.

\begin{figure}
\resizebox{\hsize}{!}{\includegraphics{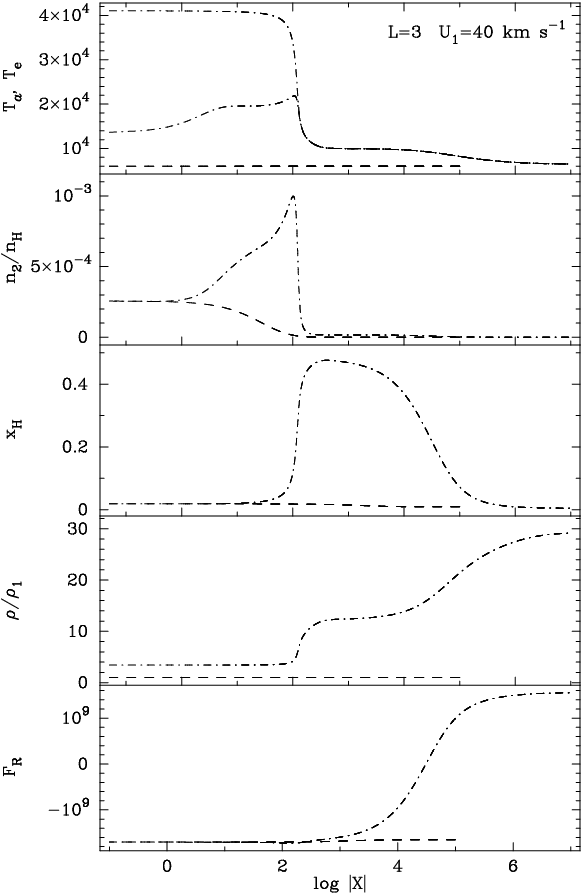}}
\caption{Same as Fig.~\ref{u20} for model with $L=3$, $U_1 = 40\kms$.}
\label{u40}
\end{figure}

Properties of shock wave models are listed also in Table~\ref{table}.
The column $U_1$ give the upstream velocity
at the outer boundary of the slab,
$\Frad$ is the radiative flux emergent from the preshock outer boundary and
$\Frad/\CCc$ is the ratio of the radiative flux to the total energy flux
at the preshock outer boundary.
Columns headed by $X_{\Jcal -1/2}$ describe the state of gas just ahead
the discontinuous jump.
Because in the preshock region the divergence of radiative flux is
negative, we use its absolute value $\dFz = |\divf|/\rho$.
The total rise of the gas temperature and the total rise of the
hydrogen ionization degree in the preshock region within the interval
$\left[X_1,X_{\Jcal -1/2}\right]$ are given in columns
$\Delta T$ and $\Delta\xH$, respectively.
The columns headed by $X_{\Jcal +1/2}$ give the temperature of heavy
particles $\Tab$, the temperature of the electron gas $\Teb$ and the
compression ratio $\eta=\rhob/\rhoa$ just behind the discontinuous jump
at the cell center $X_{\Jcal +1/2}$.
The columns headed by $\max\dFz$ give the maximum value
of the divergence of radiative flux in the postshock region
(in these layers $\divf > 0$) and
the space coordinate $X$ where this maximum is attained.
The last two columns are headed by $\max\xH$ and give the maximum ionization
degree of hydrogen $\xH$ in the postshock region as well as the distance $X$
of this maximum from the discontinuous jump.

\begin{table*}
\caption{Parameters of shock wave models with $\rho_1 = 10^{-10}\gcc$ and $T_1 = 6000\K$.}
\label{table}
\begin{tabular}{rrrrrrrrrrrrrr}
\hline
&&&&\multicolumn{3}{c}{$X_{\Jcal -1/2}$} &\multicolumn{3}{c}{$X_{\Jcal +1/2}$} &\multicolumn{2}{c}{$\max \dFz$} &\multicolumn{2}{c}{$\max \xH$} \vbox to 10pt {}\\
 $U_1$ & $M_1\;$ & $\log\Frad$ & $\Frad/\CCc$ & $\log \dFz$ & $\Delta T$ & $\Delta\xH\;$ & $\Ta^+\;$ & $\Te^+\;$ & $\rho^+/\rho^-$ & $\log \dFz$ & $\log X$   & $\xH$ & $\log X$ \vbox to 12pt {}\\[2pt]
\hline
\multicolumn{13}{l}{$L=2$} \vbox to 12pt {}\\[2pt]
 $15$& $ 1.64$& $ 7.736\;$& $ 0.176\;\;$& $ 12.35$& $   0$& $ 0.000$& $9853$& $9134$& $ 1.878$& $12.034$& $  4.49$& $  0.06$& $  4.96\;$ \\[2pt]
 $20$& $ 2.18$& $ 8.268\;$& $ 0.392\;\;$& $ 13.34$& $   0$& $ 0.000$& $13970$& $10890$& $ 2.445$& $12.759$& $  3.92$& $  0.11$& $  4.24\;$ \\[2pt]
 $25$& $ 2.73$& $ 8.593\;$& $ 0.551\;\;$& $ 14.04$& $   0$& $ 0.001$& $19113$& $12038$& $ 2.842$& $13.285$& $  3.39$& $  0.18$& $  3.74\;$ \\[2pt]
 $30$& $ 3.28$& $ 8.846\;$& $ 0.678\;\;$& $ 14.52$& $   0$& $ 0.002$& $25337$& $12801$& $ 3.116$& $13.749$& $  2.97$& $  0.26$& $  3.38\;$ \\[2pt]
 $35$& $ 3.82$& $ 9.039\;$& $ 0.731\;\;$& $ 14.89$& $   3$& $ 0.005$& $32661$& $13329$& $ 3.308$& $14.118$& $  2.67$& $  0.36$& $  3.10\;$ \\[2pt]
 $40$& $ 4.37$& $ 9.218\;$& $ 0.802\;\;$& $ 15.23$& $  14$& $ 0.010$& $41084$& $13717$& $ 3.444$& $14.387$& $  2.46$& $  0.47$& $  2.84\;$ \\[2pt]
 $45$& $ 4.91$& $ 9.369\;$& $ 0.836\;\;$& $ 15.56$& $  40$& $ 0.019$& $50582$& $14034$& $ 3.542$& $      $& $  1.89$& $  0.61$& $  2.59\;$ \\[2pt]
 $50$& $ 5.46$& $ 9.509\;$& $ 0.880\;\;$& $ 15.93$& $ 112$& $ 0.041$& $61069$& $14379$& $ 3.608$& $      $& $  1.64$& $  0.76$& $  2.40\;$ \\[2pt]
 $55$& $ 6.00$& $ 9.635\;$& $ 0.918\;\;$& $ 16.34$& $ 405$& $ 0.122$& $72054$& $15129$& $ 3.631$& $      $& $  1.26$& $  0.94$& $  2.30\;$ \\[2pt]
 $60$& $ 6.55$& $ 9.746\;$& $ 0.929\;\;$& $ 16.65$& $1617$& $ 0.366$& $81868$& $17740$& $ 3.554$& $      $& $  0.38$& $  1.00$& $  1.62\;$ \\[2pt]
 $65$& $ 7.08$& $ 9.844\;$& $ 0.918\;\;$& $ 16.65$& $3527$& $ 0.648$& $90756$& $21691$& $ 3.435$& $      $& $  0.34$& $  1.00$& $  1.24\;$ \\[2pt]
 $70$& $ 7.58$& $ 9.949\;$& $ 0.968\;\;$& $ 16.33$& $5029$& $ 0.862$& $100569$& $24765$& $ 3.365$& $      $& $  0.30$& $  1.00$& $  1.06\;$ \\[2pt]
\hline
\multicolumn{13}{l}{$L=3$} \vbox to 12pt {}\\[2pt]
 $15$& $ 1.64$& $ 7.794\;$& $ 0.199\;\;$& $ 12.34$& $   0$& $ 0.000$& $9866$& $9166$& $ 1.888$& $12.196$& $  4.46$& $  0.05$& $  4.93\;$ \\[2pt]
 $20$& $ 2.18$& $ 8.272\;$& $ 0.386\;\;$& $ 13.24$& $   0$& $ 0.000$& $13900$& $10739$& $ 2.395$& $12.900$& $  3.79$& $  0.11$& $  4.17\;$ \\[2pt]
 $25$& $ 2.73$& $ 8.593\;$& $ 0.539\;\;$& $ 13.91$& $   0$& $ 0.001$& $19051$& $11913$& $ 2.798$& $13.491$& $  3.19$& $  0.18$& $  3.62\;$ \\[2pt]
 $30$& $ 3.28$& $ 8.843\;$& $ 0.657\;\;$& $ 14.35$& $   2$& $ 0.002$& $25283$& $12704$& $ 3.079$& $13.961$& $  2.79$& $  0.26$& $  3.22\;$ \\[2pt]
 $35$& $ 3.82$& $ 9.039\;$& $ 0.720\;\;$& $ 14.68$& $   7$& $ 0.005$& $32614$& $13253$& $ 3.277$& $14.291$& $  2.52$& $  0.36$& $  2.90\;$ \\[2pt]
 $40$& $ 4.37$& $ 9.218\;$& $ 0.790\;\;$& $ 14.98$& $  16$& $ 0.009$& $41041$& $13654$& $ 3.419$& $14.525$& $  2.32$& $  0.48$& $  2.63\;$ \\[2pt]
 $45$& $ 4.91$& $ 9.368\;$& $ 0.826\;\;$& $ 15.29$& $  39$& $ 0.018$& $50541$& $13976$& $ 3.521$& $14.701$& $  2.17$& $  0.61$& $  2.41\;$ \\[2pt]
 $50$& $ 5.46$& $ 9.513\;$& $ 0.887\;\;$& $ 15.62$& $ 102$& $ 0.041$& $61026$& $14310$& $ 3.591$& $14.861$& $  2.09$& $  0.77$& $  2.24\;$ \\[2pt]
 $55$& $ 6.00$& $ 9.630\;$& $ 0.891\;\;$& $ 16.03$& $ 327$& $ 0.117$& $72040$& $14919$& $ 3.621$& $      $& $  2.26$& $  0.95$& $  2.18\;$ \\[2pt]
 $60$& $ 6.55$& $ 9.744\;$& $ 0.916\;\;$& $ 16.43$& $1203$& $ 0.321$& $82504$& $16849$& $ 3.578$& $      $& $  3.46$& $  1.00$& $  1.57\;$ \\[2pt]
 $65$& $ 7.08$& $ 9.846\;$& $ 0.924\;\;$& $ 16.59$& $2651$& $ 0.541$& $92943$& $19958$& $ 3.504$& $      $& $  4.11$& $  1.00$& $  1.20\;$ \\[2pt]
 $70$& $ 7.58$& $ 9.970\;$& $      $& $ 16.51$& $4107$& $ 0.748$& $103543$& $23023$& $ 3.438$& $      $& $  4.55$& $  1.00$& $  1.02\;$ \\[2pt]
\hline
\multicolumn{13}{l}{$L=4$} \vbox to 12pt {}\\[2pt]
 $15$& $ 1.64$& $ 7.755\;$& $ 0.176\;\;$& $ 12.26$& $   0$& $ 0.000$& $9716$& $8886$& $ 1.803$& $12.240$& $  4.49$& $  0.05$& $  4.97\;$ \\[2pt]
 $20$& $ 2.18$& $ 8.276\;$& $ 0.385\;\;$& $ 13.14$& $   0$& $ 0.000$& $13866$& $10667$& $ 2.371$& $12.967$& $  3.77$& $  0.11$& $  4.16\;$ \\[2pt]
 $25$& $ 2.73$& $ 8.587\;$& $ 0.521\;\;$& $ 13.79$& $   0$& $ 0.001$& $19023$& $11854$& $ 2.777$& $13.556$& $  3.16$& $  0.18$& $  3.59\;$ \\[2pt]
 $30$& $ 3.28$& $ 8.835\;$& $ 0.633\;\;$& $ 14.24$& $   3$& $ 0.002$& $25347$& $12827$& $ 3.124$& $14.020$& $  2.75$& $  0.26$& $  3.18\;$ \\[2pt]
 $35$& $ 3.82$& $ 9.040\;$& $ 0.717\;\;$& $ 14.55$& $   7$& $ 0.005$& $32592$& $13215$& $ 3.263$& $14.329$& $  2.51$& $  0.36$& $  2.86\;$ \\[2pt]
 $40$& $ 4.37$& $ 9.210\;$& $ 0.761\;\;$& $ 14.85$& $  16$& $ 0.009$& $41022$& $13622$& $ 3.407$& $14.557$& $  2.30$& $  0.48$& $  2.59\;$ \\[2pt]
 $45$& $ 4.91$& $ 9.372\;$& $ 0.835\;\;$& $ 15.16$& $  38$& $ 0.018$& $50524$& $13948$& $ 3.510$& $14.729$& $  2.14$& $  0.61$& $  2.38\;$ \\[2pt]
 $50$& $ 5.46$& $ 9.504\;$& $ 0.851\;\;$& $ 15.51$& $  99$& $ 0.040$& $61089$& $14365$& $ 3.614$& $14.889$& $  2.05$& $  0.77$& $  2.21\;$ \\[2pt]
 $55$& $ 6.00$& $ 9.630\;$& $ 0.884\;\;$& $ 15.94$& $ 297$& $ 0.113$& $72061$& $14837$& $ 3.616$& $      $& $  2.23$& $  0.95$& $  2.16\;$ \\[2pt]
 $60$& $ 6.55$& $ 9.743\;$& $ 0.909\;\;$& $ 16.38$& $1069$& $ 0.302$& $82746$& $16553$& $ 3.584$& $      $& $  3.41$& $  1.00$& $  1.56\;$ \\[2pt]
 $65$& $ 7.08$& $ 9.847\;$& $ 0.923\;\;$& $ 16.58$& $2377$& $ 0.503$& $93630$& $19398$& $ 3.524$& $      $& $  4.04$& $  1.00$& $  1.19\;$ \\[2pt]
 $70$& $ 7.58$& $ 9.973\;$& $      $& $ 16.57$& $3772$& $ 0.696$& $104720$& $22378$& $ 3.466$& $      $& $  4.47$& $  1.00$& $  1.01\;$ \\[2pt]
\hline
\end{tabular}
\end{table*}

A cursory inspection of Table~\ref{table} shows that the total radiative flux
$\Frad$ nearly does not depend on the number of bound atomic levels $L$.
In other words, the radiative flux emergent from the shock wave
is independent of the lower limit $\nu_L$ of the frequency range
provided that $\nu_L < \nu_2$.
At the same time, the change of $\nu_L$ is accompanied by redistribution
of the energy of radiation field within $[\nu_L ,\nu_0]$ which is
most perceptible within the Lyman continuum
In particular, for the narrower frequency range (e.g. for $L=2$)
we have both stronger radiative heating and higher hydrogen ionization
in the preshock region.
The structure of the postshock region is also sensitive to the width of
the frequency range $[\nu_L,\nu_0]$.
In particular, for the narrower frequency range the thermal equipartition
and the maximum of ionization
are attained at larger distances from the discontinuous jump.


\section{Radiation field}\label{radfld}

\begin{figure}
\resizebox{\hsize}{!}{\includegraphics{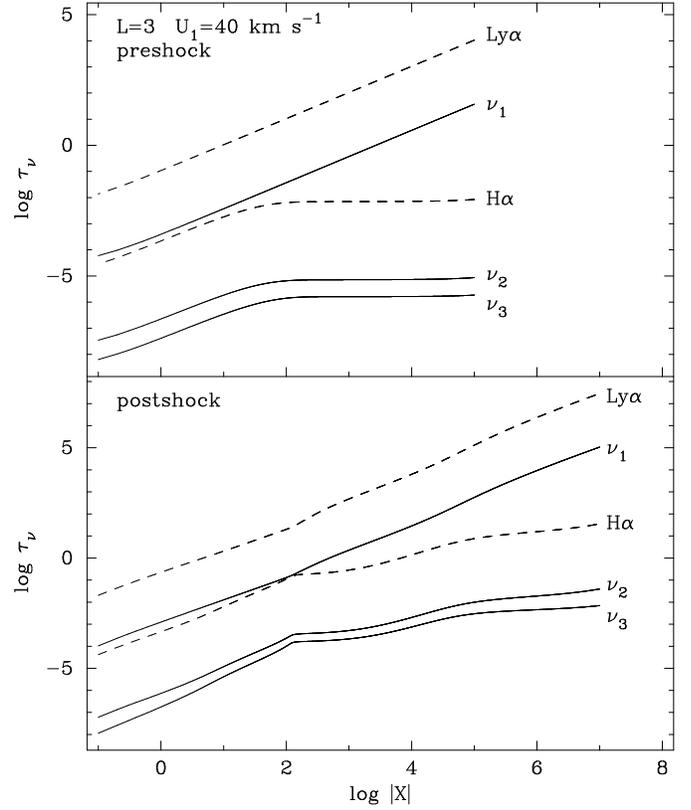}}
\caption{Monocromatic optical depths $\tau_\nu$ measured from the
discontinuous jump for the preshock (upper panel) and
postshock (lower panel) regions of the model with $L=3$, $U_1=40\kms$.
In solid lines are shown the optical depths at the heads of the continua
and in dashed lines are shown the optical depths at the centers of
spectral lines.}
\label{u40tau}
\end{figure}

We consider the shock waves propagating through the partially ionized
hydrogen gas which is opaque both in the Lyman continuum and in the
Lyman lines.
The total optical depth of the slab at the frequency $\nu_1$
corresponding to the threshold ionization from the ground state
is in the range $10^4\lesssim\tau(\nu_1)\lesssim 10^5$,
the maximum $\tau(\nu_1)\sim 10^5$ being reached at $U_1 \approx 40\kms$.
At the same time the slab is nearly transparent for
continuum radiation at frequencies $\nu < \nu_1$.
The optical depths in the centers of Balmer lines are
$\tau_\nu\ll 1$ in the preshock region and gradually increase
with increasing distance from the discontinuous jump in the postshock region.
At distances $X\gtrsim 10^4\cm$ the postshock optical depth in the center
of the \Ha\ spectral
line measured from the discontinuous jump is $\tau_\nu\gtrsim 1$.
In Fig.~\ref{u40tau} we give the plots of monochromatic optical depths
$\tau_\nu$ measured from the discontinuous jump to the layer with
space coordinate $X$
for the heads of continua with threshold frequencies $\nu_i$
$(1\le i\le 3)$ and for the centers of spectral lines \Lya\ and \Ha.

Thus, the Eddington approximation cannot be applied for the solution of the
radiation transfer problem in shock waves. Indeed, it is valid only
within the Lyman continuum whereas the most of radiation is transported
at frequencies lower than the Lyman edge where
the angular dependence of the specific intensity $I_{\mu\nu}$
must be taken into account.
Moreover, the correct calculation of the angular dependence of the
specific intensity $I_{\mu\nu}$ is of great importance because
for $\nu < \nu_1$ the term
\begin{equation}
\frac{1}{\mu}\int\limits_0^{\tau_\nu} S_\nu(t)
e^{-(t - \tau_\nu)/\mu}dt
\end{equation}
rapidly decreases for $\mu\to |1|$ and the Eddington factor is
$f_\nu = K_\nu/J_\nu < 1/3$.

\begin{figure}
\resizebox{\hsize}{!}{\includegraphics{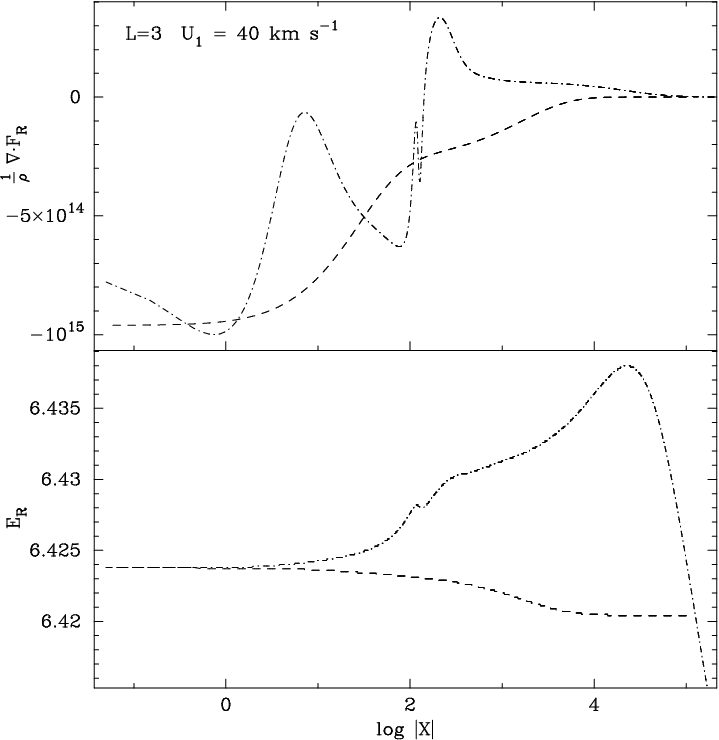}}
\caption{The divergence of radiative flux per unit mass (upper panel) and
the radiation energy density (lower panel) as a function of the
distance from the discontinuous jump for the model with
$L=3$, $U_1=40\kms$. In dashed and dot--dashed lines are given the
preshock and postshock plots, respectively.}
\label{u40erad}
\end{figure}

The interaction between the radiation field and the gas material is described
in terms of the divergence of radiative flux $\divf$.
By definition, this quantity is negative when the gas absorbs more energy
than it emits, whereas when $\divf > 0$, the gas radiatively cools.
The typical plot of the divergence of radiative flux
expressed per unit mass $\divF$
is shown on the upper panel of Fig.~\ref{u40erad} for the
shock wave model with $L=3$ and $U_1 = 40\kms$.
The gradual decrease of $\divF$ in the preshock region is due to the
strong absorption of the Lyman continuum radiation which leads to the
growth of the gas temperature when the parcel of gas approaches the
discontinuous jump.
The heating of gas rapidly increases with increasing upstream velocity
and as is seen from Table~\ref{table} the quantity $\dFz = |\divf|/\rho$
at $X_{\Jcal -1/2}$ changes be several orders of magnitude for
$15\kms\le U_1\le 70\kms$.

The jump of $\divF$ within $[X_{\Jcal -1/2}, X_{\Jcal +1/2}]$
is due to the adiabatic heating of electrons at the discontinuous jump.
The layers of the most efficient radiative cooling in the postshock region
with fastest decrease of the gas temperature are revealed by the maximum
of $\divF$ whicj also very rapidly increases with increasing $U_1$
(see Table~\ref{table}).
The divergence of radiative flux approaches zero with increasing distance
from the discontinuous jump in the both preshock and postshock regions.
This implies that at large distances from the discontinuous jump
the total radiative flux $\Frad$ tends to be constant.

While the parcel of gas approaches the
discontinuous jump in the preshock region the radiation energy density
$\Erad$ gradually increases due to the radiative heating of gas.
Behind the discontinuous jump $\Erad$ continues to increase and reaches
the maximum in the layers of the fastest recombination of hydrogen atoms.
The maximum of the radiation energy density $\Erad$ in the postshock region
approximately corresponds to the point where the total radiative flux is
$\Frad = 0$.
It should be noted, however, that this condition is exact only in the
diffusion approximation (see, e.g., Mihalas \& Mihalas 1984) and as is
shown below the space coordinate of the layer with $F_\nu = 0$
depends on the frequency $\nu$.
The typical plot of $\Erad$ is shown on the lower panel of Fig.~\ref{u40erad}.
For all models considered in the present study the radiation energy
density $\Erad$ can be neglected in comparison with the thermal energy
of the gas. For example, in the model represented in Fig.~\ref{u40erad}
the ratio of the radiation energy density to the enthalpty
is $\Erad/h\lesssim 10^{-3}$ throughout the whole shock wave.

\begin{figure}
\resizebox{\hsize}{!}{\includegraphics{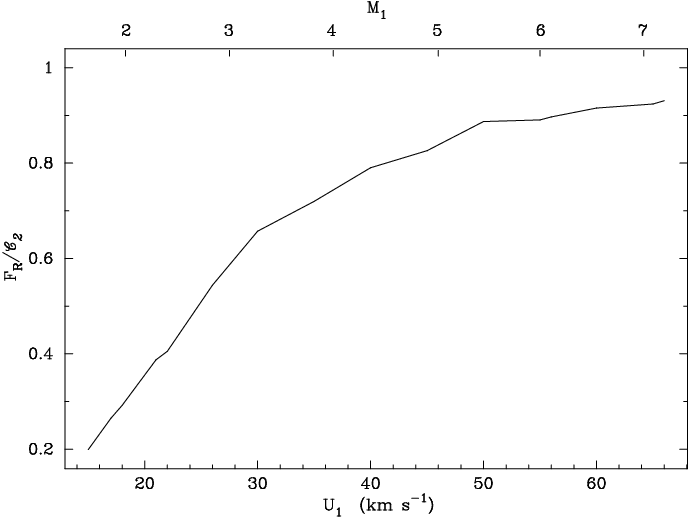}}
\caption{The ratio of the radiative flux $\Frad$ emergent from the preshock
outer boundary to the total energy of the shock wave $\CCc$ versus the
upstream gas flow velocity $U_1$ for models with $L=3$.
Along the upper horizontal axis are given the upstream adiabatic Mach numbers
$M_1$.}
\label{frc2}
\end{figure}

The radiative flux outgoing from the preshock outer boundary is nearly the
same as that outgoing from the postshock outer boundary, that is,
$-F_1 \approx F_N$.
The radiative flux emergent from the shock wave rapidly increases with
increasing upstream velocity and at $U_1\approx 60\kms$ more than
90 percent of the total energy flux is contained in the radiation flux.
As is seen from Fig.~\ref{frc2} where we give the plot of the ratio
$\Frad/\CCc$ versus the upstream velocity $U_1$, at upstream Mach numbers
$M_1 \gtrsim 6$ the major part of the shock wave energy
(i.e. more than 90\%) is lost due to radiation.

At upstream velocities $U_1 < 40\kms$ the total radiative flux is
nearly constant in the preshock region and undergoes the substantial
changes only behind the discontinuous jump in the layers where the gas
radiatively cools and recombines.
However the role of the Lyman continuum radiation rapidly increases
with increasing upstream velocity and for $U_1 > 40\kms$
the total radiative flux $\Frad$ shows perceptible changes
in the preshock region (see Fig.~\ref{u60}).

\begin{figure}
\resizebox{\hsize}{!}{\includegraphics{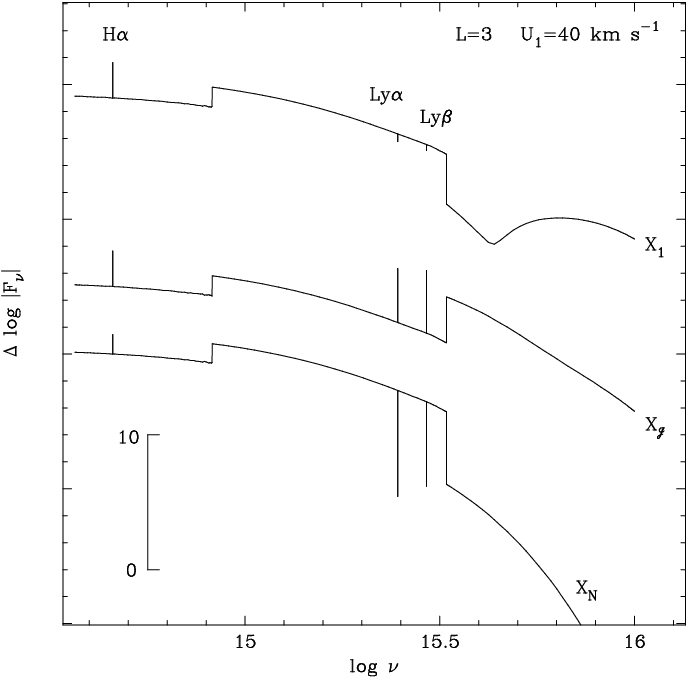}}
\caption{Monochromatic radiative flux $F_\nu$ as a function of frequency $\nu$
for the preshock outer boundary $X_1$, discontinuous jump $X_\Jcal$ and
postshock outer boundary $X_N$.
For the sake of convenience the plots are arbitrarily shifted along the
vertical axis.}
\label{u40fnu}
\end{figure}

Of great interest are the spectral distributions of monocromatic
quantities describing the radiation field as a function of frequency $\nu$.
In Fig.~\ref{u40fnu} we give the spectra of monocromatic radiative flux $F_\nu$
for three different layers:
the preshock outer boundary $X_1$, the discontinuous jump
$X_\Jcal$ and the postshock outer boundary $X_N$
in the model with $L=3$ and $U_1 = 40\kms$.
As is seen from these plots, the most of radiation is transported within
the frequency interval $\nu_2 < \nu < \nu_1$ limited by the edges of
the Balmer and  Lyman continua.
The contribution of the Lyman continuum radiation is perceptible only
in the vicinity of the discontinuous jump.

The conspicuous feature of these plots is also that the \Ha\ spectral
line is in emission throughout the shock wave, whereas the
\Lya\ and \Lyb\ lines are mostly in absoprtion and are revealed
in emission only in the vicinity of the discontinuous jump.
Within the range of upstream velocities $15\kms\le U_1\le 70\kms$
the frequency dependencies of the radiative flux are qualitatively similar
to that shown in Fig.~\ref{u40fnu}.

\begin{figure}
\resizebox{\hsize}{!}{\includegraphics{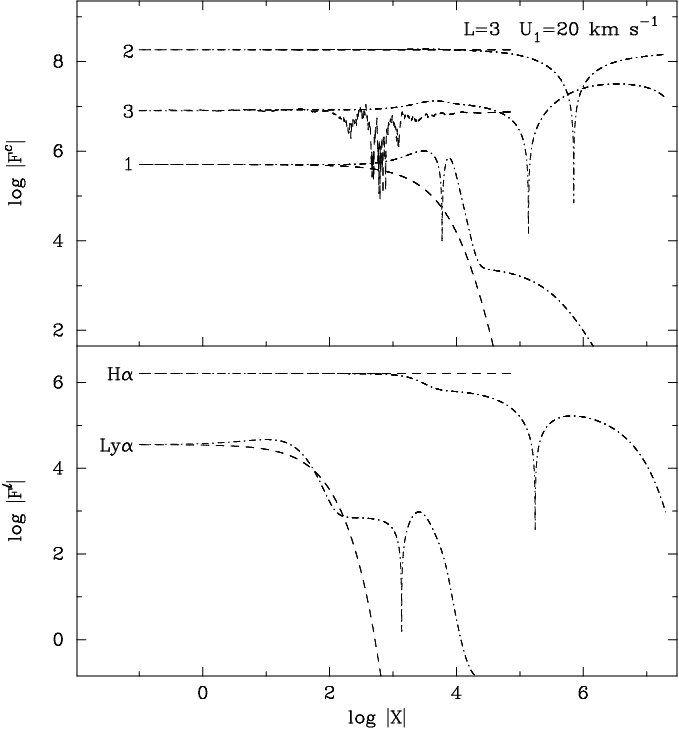}}
\caption{The radiative flux in continuum intervals $F^c_i$ (upper panel)
and in spectral lines (lower panel) for model with $L=3$, $U_1=20\kms$.
In dashed and dot--dashed lines are represented the preshock and
postshock plots, respectively.}
\label{u20frad}
\end{figure}

\begin{figure}
\resizebox{\hsize}{!}{\includegraphics{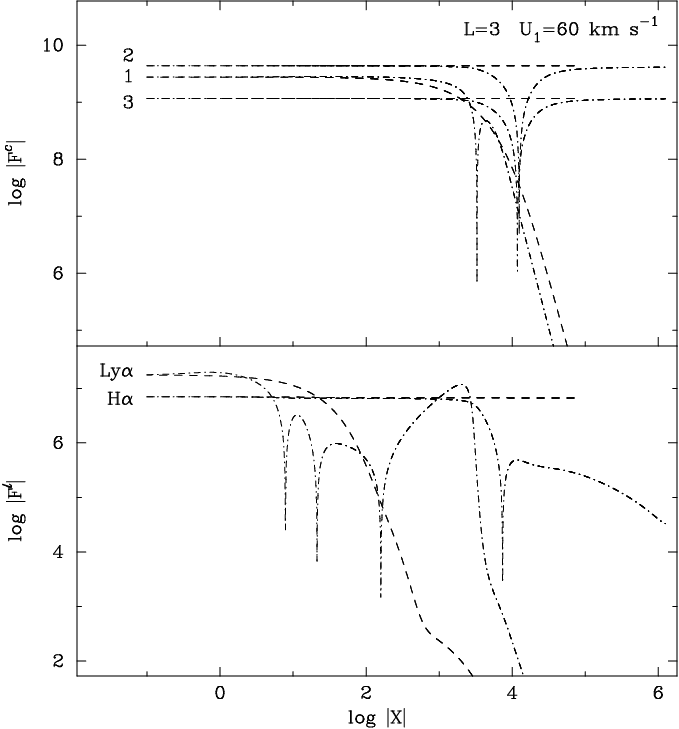}}
\caption{Same as Fig.~\ref{u20frad} for model with $L=3$, $U_1 = 60\kms$.}
\label{u60frad}
\end{figure}

The role of various spectral intervals in the total radiation transfer
can be evaluated with use of terms
\begin{equation}
F^c_i = 4\pi\int\limits_{\nu_i}^{\nu_{i-1}} H_\nu d\nu
\end{equation}
and
\begin{equation}
F^\ell_i =4\pi\int\limits_{\nu_i-\Delta\nu_i}^{\nu_i+\Delta\nu_i} H_\nu d\nu
\end{equation}
representing the radiative flux integrated within the continuum
frequency intervals $[\nu_i,\nu_{i-1}]$
and within the spectral lines, respectively.
Here $\Delta\nu_i$ is the half--width of the spectral line
frequency interval.
Figs.~\ref{u20frad}\,--\,\ref{u60frad} show $F^c_i$ and $F^\ell_i$
as a function of the distance from the discontinuous jump
for models with $U_1 = 20$, 40 and $60\kms$, respectively.
For the sake of convenience of the graphical representation the fluxes
are given in the logarithmic scale.
As is seen from these plots throughout the shock wave the most of radiation
is transported within the Balmer continuum.
The role of the Lyman continuum becomes perceptible at upstream velocities
$U_1 > 40\kms$ only in the vicinity of the
discontinuous jump: $-10^4~\cm \lesssim X \lesssim 10^4~\cm$.
In the preshock region the radiative flux is constant at frequencies
$\nu < \nu_1$.
The Lyman continuum flux ($\nu_1\le\nu\le\nu_0$) rapidly increases
with approaching the discontinuous jump and becomes perceptible at
distances smaller than $X\sim 10^4\cm$.
But even in the vicinity of the discontinuous jump the radiative energy
transport in the Lyman continuum remains small in comparison with
that at lower frequencies until the upstream velocity is
$U_1\lesssim 40\kms$.

\begin{figure}
\resizebox{\hsize}{!}{\includegraphics{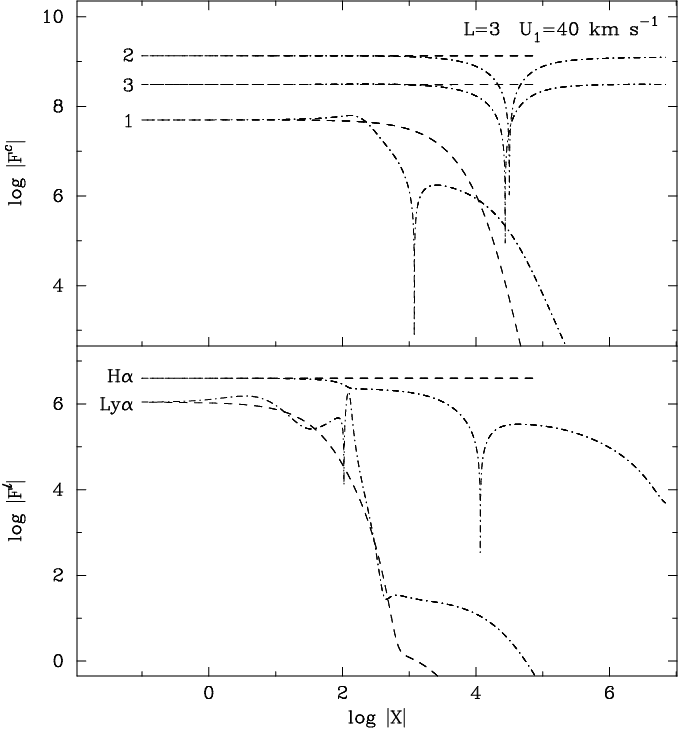}}
\caption{Same as Fig.~\ref{u20frad} for model with $L=3$, $U_1 = 40\kms$.}
\label{u40frad}
\end{figure}

As is seen from the plots given on the lower panels of
Figs.~\ref{u20frad}\,--\,\ref{u60frad} the contribution of the spectral line
radiation into the total radiative flux is negligible.
Throughout the slab the most of the spectral line radiation is
transported by \Ha\ and only at upstream velocities $U_1 > 50\kms$
in the close vicinity of the discontinuous jump the spectral line
radiation is transported mostly in \Lya.

Behind the discontinuous jump the monochromatic radiative flux undergoes
substantial changes within the radiative cooling region from where the
most of radiation is emerged in both opposite directions.
By definition, the monochromatic radiative flux changes its sign in
the layer where $\divfnu$ reaches the maximum.
In plots given in Figs.~\ref{u20frad}\,--\,\ref{u60frad} these layers are
revealed as deep maximuma of the continuum and spectral line fluxes
$F^c_i$ and $F^\ell_i$, respectively.

As is seen, the region emerging radiation is spread along the space coordinate
in wide ranges:
$3.8 \lesssim \log X \lesssim 6.5$ for $U_1 = 20\kms$ and
$3.5 \lesssim \log X \lesssim 4.1$ for $U_1 = 60\kms$.
This is obviously due to strong frequency dependence of optical depths
$\tau_\nu$ in layers emitting the radiation.


\section{Preshock region}\label{preshreg}

Effects of the electron thermal conduction in the vicinity of the
discontinuous jump are not considered in the present study,
therefore throughout the preshock region the translational energies
of the both hydrogen atoms and free electrons are equal and can be
described in terms of the gas temperature: $T=\Ta=\Te$.
For $T=6000\K$ and $\rho=10^{-10}\gcc$
most of the hydrogen atoms are in the ground state, so that
the zone in which the radiation field interacts with gas material
is only of several mean free paths of photons at the frequency of the Lyman
continuum edge $\nu_1$. For our models this corresponds to the
distance from the discontinuous jump of $X\sim 10^4\cm$.
Within this zone known as the radiative precursor
the gas undergoes both radiative heating and photoionization.
The rize of the gas temperature $\Delta T$ and the hydrogen
ionization degree $\Delta\xH$ within the preshock region is nearly
negligible in shocks with upstream velocities smaller than $U_1 \approx 30\kms$
but at larger upstream velocities $\Delta T$ and $\Delta\xH$
very rapidly grow with increasing $U_1$ (see Table~\ref{table}).
At upstream velocities $U_1 > 65\kms$ the hydrogen ionization degree
ahead the discontinuous jump is $\xH > 0.5$ and the gas material
becomes more transparent. This accounts for the small decrease of
$\dFz=|\divf|/\rho$ for $U_1 > 65\kms$.

\begin{figure}
\resizebox{\hsize}{!}{\includegraphics{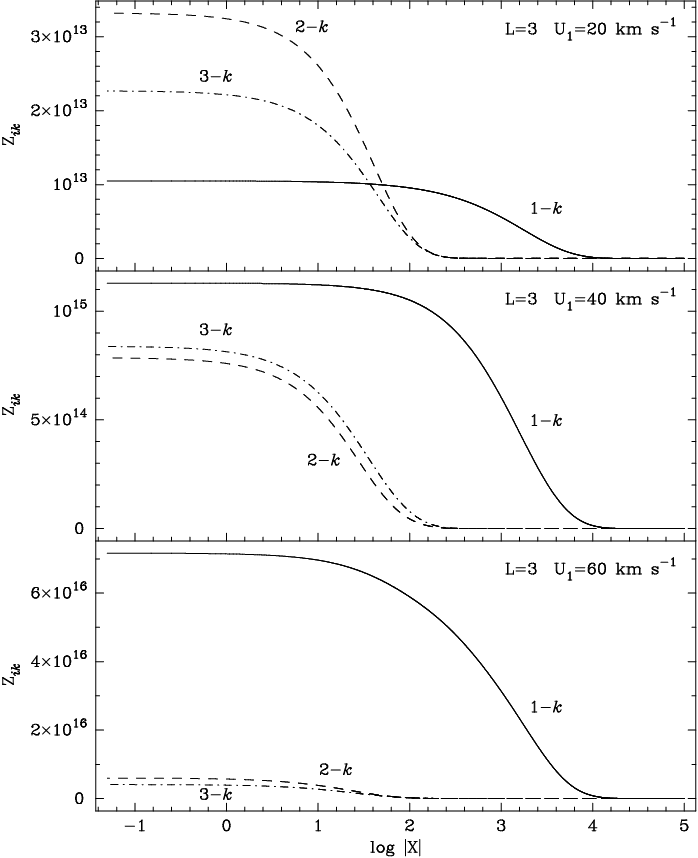}}
\caption{Net radiative ionization  rates for the ground state
(solid line), second level (dashed line) and third level
(dot--dashed line) in the preshock regions of the
shock wave models with $U_1 = 20\kms$ (upper panel),
$U_1 = 40\kms$ (middle panel) and $U_1 = 60\kms$ (lower panel).}
\label{inrprc}
\end{figure}

Throughout the preshock region the collisional ionizations can be
neglected since their rates are by several orders of magnitude smaller
than those of photoionizations.
In Fig.~\ref{inrprc} we give the plots of the net rates of
radiative ionizations
\begin{equation}
\label{nphir}
\Zik = \nni\Rik - \nel\frac{\nnilte}{\nklte}\Rikd
\end{equation}
for first three atomic levels in models with $U_1 = 20$, 40 and $60\kms$.
At distances from the discontinuous jump less than $X \sim 10^2\cm$
the absorption coefficient in \Lya\ and \Lyb\ lines exceeds that
of the Lyman continuum.
For upstream velocities $U_1 < 30\kms$ the ionization from the ground
state is still small and in layers with $X < 10^2\cm$
the hydrogen ionizes mostly from excited levels $i \ge 2$.
However the rapid growth of the ionization from the ground state
with increasing upstream velocity cancels this effect and
at $U_1 > 30\kms$ the hydrogen ionizes
mostly from the ground state.

The columns $\Delta T$ and $\Delta\xH$ of Table~\ref{table} give
the total growth of the gas temperature and ionization degree
in the preshock region. As is seen, at constant upstream velocity
$U_1$ both $\Delta T$ and $\Delta\xH$ decrease for larger $L$.
As was noted above the total radiative flux emitted by the shock wave
nearly does not depend on the frequency range $[\nu_L,\nu_0]$ for $L>2$
and for smaller $\nu_L$ the monocromatic flux is redistributed among
lower frequencies. In particular, the fraction of the Lyman continuum
radiation in the total radiative flux decreases.

The heating of the gas within the radiative precursor is accompanied
by the rise of the gas pressure $\Pgas$ and the gas density $\rho$.
However the increase of $\rho$ is not large and for $U_1 = 60\kms$
the gas density just ahead the discontinuous jump is $\rhoa = 1.0085\rho_1$,
so that the upstream gas flow velocity decreases by less than one
per cent and is $\Ua = 0.9916 U_1$.
More important is the rise of the gas temperature $T$
because it leads to the slow decrease of the compression ratio
$\rhob/\rhoa$ with increasing $U_1$ in shocks with
upstream velocities $U_1 > 55\kms$ (see Table~\ref{table}).
Such a decrease of the compression ratio $\rhob/\rho_1$ with
increasing upstream velocity is easily explained in terms of the
Rankine--Hugoniot relations (\ref{C0}) -- (\ref{C2}).
Adopting $\Frada = \Fradb$ in (\ref{Aterm}) and neglecting the contribution
of the radiation energy density $\Erad$ and radiation pressure $\Prad$
in Eqs.~(\ref{Aterm}) and (\ref{Bterm})
we find that the compression ratio at the discontinuous jump is
\begin{equation}
{\rhob\over\rhoa} = {4\left(\Ua\right)^2\over 5\left(\aTa\right)^2 + \left(\Ua\right)^2}.
\end{equation}
Thus, the slow decrease of the compression ratio $\rhob/\rhoa$ with increasing
$U_1$ is due to the rize of the isothermal sound speed ahead the
discontinuity jump.
Inspection of Table~\ref{table} shows that the effect of decreasing
$\rhob/\rhoa$
becomes less prominent in the models computed with larger $L$ and,
therefore, having somewhat weaker radiative heating of the gas just ahead the
discontinuous jump.

For gas densities of $10^{-14}\gcc\lesssim\rho\lesssim 10^{-8}\gcc$
the dominant recombination mechanism is the three--body recombination
(see, e.g., Zeldovich \& Raizer 1967) and as was shown by
Gillet \& Lafon (1990) the precursor heating is due to the
partial transformation of the ionization energy into the thermal energy
via the three--body recombination processes.
Kuznetsov \& Raiser (1965) have shown that electrons are captured
on the level with energy $U_i = \chiH/i^2$ close to the average
kinetic energy of free electrons $k\Te$.
Thus, $i=\sqrt{\chiH/ k\Te}$ and for the precursor temperature range
$6000\K\le\Te\lesssim 10^4\K$ (see Table~\ref{table})
the free electrons should be captured onto levels $4\le i\le 5$.
Thus, the atomic model represented by 4 bound levels seems do not
underestimate significantly the rates of three--body recombinations
and, therefore, the precursor heating is treated quite correctly.
This conclusion is consistent with decrease of the temperature rize
$\Delta T$ obtained for atomic models with $L=3$ and $L=4$
(see Table~\ref{table}).


\section{Postshock region}\label{postshreg}

Just behind the discontinuous jump the temperature of neutral hydrogen atoms
and hydrogen ions $\Tab$ exceeds the temperature of free electrons $\Teb$
(see Table~\ref{table}) and as is seen from Figs.~\ref{u20}\,--\,\ref{u60}
both these temperatures tend to equalize with increasing distance $X$.
Together with excitation of bound atomic levels the thermal equilibration
is the fastest relaxation process behind the discontinuous jump.
For the models considered in the present study the hydrogen ionization
degree ahead the discontinuous jump is $\xH^- > 10^{-2}$, so that
ionized hydrogen atoms are enough numerous behind the discontinous jump and
free electrons gain the energy from heavy particles mostly in
elastic collisions with hydrogen ions.
It should be noted that for all models throughout the postshock region
$\Qei\gg\Qea$.

When the temperatures of heavy particles and free electrons equalize,
the gas radiatively cools and the gas temperature $T=\Ta=\Te$
gradually decreases with increasing distance from the discontinuos jump.
The radiative cooling is fastest when the divergence of radiative flux
(\ref{divf}) reaches its maximum.
The maximum postshock values of $\divF$ and the space coordinate $X$,
where this maximum is reached, are given in Table~\ref{table}.
The maximum of $\divF$ is prominent only when the postshock ionization
degree is $\xH < 0.8$ and for larger $\xH$ the spatial dependence
of $\divF$ shows the plateau rather than the maximum.
It should be noted that the maximum of $\divf$ gives only the
frequency averaged location of the layers emitting the radiation.
As was shown above the space coordinate of these layers depends on the
frequency of radiation $\nu$.

\begin{figure}
\resizebox{\hsize}{!}{\includegraphics{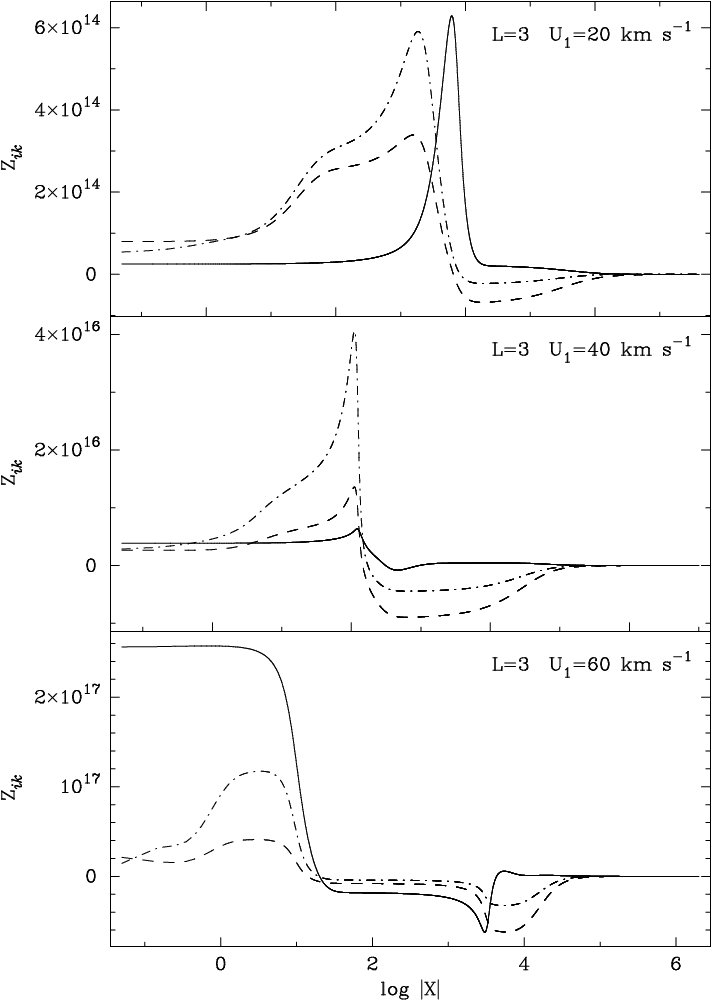}}
\caption{Net radiative ionization rates from the ground state
(solid line), from the second level (dashed line) and from the third level
(dot--dashed line) in the postshock region of the
shock wave models with $U_1 = 20\kms$ (upper panel),
$U_1 = 40\kms$ (middle panel) and $U_1 = 60\kms$ (lower panel).}
\label{inrwak}
\end{figure}

The ionization of hydrogen atoms is the slower relaxation process in
comparison with thermal equilibration and
the maximum of $\xH$ is reached behind the maximum of $\divF$
(see Table~\ref{table}).
Throughout the postshock region the rates of collisional processes
are by several orders of magnitude smaller than those of radiative processes
and, therefore, can be neglected.
Fig.~\ref{inrwak} shows the net photoionization rates $\Zik$
for the first three bound levels of the hydrogen atom in models
with $U_1 = 20$, 40 and $60\kms$.
In the postshock region the bound atomic levels are excited faster
than hydrogen ionizes, so that at upstream velocities $U_1 < 50\kms$
hydrogen atoms ionize mostly from bound levels $i\ge 2$.
At upstream velocities $U_1 > 55\kms$ more than ten percent of hydrogen
ionizes in the precursor and in the postshock region the
hydrogen ionizes mostly from the ground state.
At $U_1 > 55\kms$ the maximum postshock ionization degree is $\xH\approx 1$.
This leads to the slower postshock temperature decrease
until the hydrogen ionization degree is near the maximum
(see the upper panel of Fig.~\ref{u60}).

The full thickness of the relaxation zone behind the discontinuous jump
is determined by the slowest relaxation process which in our case is
the recombination of hydrogen atoms.
Substitution of the translational energies of hydrogen atoms and free electrons
$\Ett = \Eat + \Eet$ into the Rankine--Hugoniot equations
(\ref{C1}) and (\ref{C2}), where we neglect terms with $\Erad$ and $\Prad$,
gives
\begin{equation}
\label{rhe}
\begin{array}{l}
\left(\Ett - \widetilde{E}_\mathrm{t1}\right) +
\left(\Eext - \widetilde{E}_\mathrm{ex1}\right) +
\left(\Eit - \widetilde{E}_\mathrm{I1}\right) = \\[8pt]
= {\dst\frac{1}{2}\left(\Pgas + P_\mathrm{g1}\right){\compr -1\over\rho} +
{F_\mathrm{R1} - \Frad\over\dot m}} .
\end{array}
\end{equation}
Here subscript 1 denotes the quantities defined at the preshock outer
boundary $X_1$.
For strong shock waves with $\widetilde{E}_\mathrm{t1}\ll\Ett$,
$\widetilde{E}_\mathrm{ex1}\ll\Eext$ and $\widetilde{E}_\mathrm{I1}\ll\Ei$
the relation (\ref{rhe}) is rewritten as
\begin{equation}
\label{comprx}
\compr = 4 + 3{\Eext + \Eit\over\Ett} + 3{\Frad - F_\mathrm{R1}\over\dot m\Ett} .
\end{equation}
The first term in (\ref{comprx}) is the limiting compression ratio
$(\gamma +1)/(\gamma -1) = 4$ for the strong adiabatic shock front ($M_1\gg 1$).
In fact, because of radiative processes in the precursor
the compression ratio at the adiabatic shock front of our models
is as large as $\rhob/\rhoa =3.6$ (see Table~\ref{table}).
The second and the third terms in (\ref{comprx}) describe the additional
gas compression which occurs behind the discontinuous jump in the
postshock region due to excitation, ionization and recombination of
hydrogen atoms.
It should be noted that because $F_\mathrm{R1} < 0$, the third term in
(\ref{comprx}) is always positive.
Thus, the transfer of the part of the gas flow kinetic energy
into internal degrees of freedom of atoms and into radiation field
produces the gas density increase behind the discontinuous jump
which is much larger than that at the adiabatic shock front.
For example, in the model with $U_1 = 60\kms$ we obtained the final
compression ratio as high as $\rho/\rho_1\approx 65$ (see Fig.~\ref{u60})
which approaches the maximum compression ratio of $\gamma M_1^2 \approx 84$
in the isothermal shock wave where all the shock energy is transformed
into the radiation.


\section{Conclusion}\label{conclus}

In the work presented here we have substantially improved the computer code
for calculating the structure of radiative shock wave models and
considered almost the entire zone of postshock hydrogen recombination
in shock wave models with upstream Mach numbers $1.6\le M_1\le 7.5$.
This allowed us to obtain the reliable estimates of the radiative flux
emitted by the shock wave as a function of the upstream gas flow velocity.
The most striking result of our calculations is that the ratio
of the radiative flux to the total energy flux of the shock wave
very rapidly enlarges with increasing upstream velocity, so that
for adiabatic Mach numbers $M_1 > 7$ (i.e. at $U_1 > 65\kms$)
the major part of the shock wave energy (more than 90\%)
is irreversibly lost due to dissipation processes.

Another noteworthy feature of our results is the negligible role of collisional
processes in the both bound--bound and bound--free transitions
in comparison with radiative transitions
throughout the whole shock wave. This circumstance makes our models
independent of uncertainties in cross sections of collisional processes.

The results described in the present paper were obtained for the
only case with gas temperature $T_1 = 6000\K$ and gas density
$\rho_1 = 10^{-10}\gcc$ of the ambient unperturbed medium
which is most typical for atmospheres of classical Cepheids and
RR~Lyr type variables.
Obviously, many more models have to be constructed for various gas
temperatures and gas densities in order to more clearly understand
the properties of radiative shock waves observed in astrophysical
phenomena.
Of particular interest are the gas temperatures and gas densities
typical for pulsating late--type stars possessing strong
shock--driven stellar winds. 
The efforts of further calculations will be also directed to the
more accurate modelling of strong shock waves
in order to consider in detail the run of the ratio $\Frad/\CCc$
as a function of the upstream velocity at Mach numbers $M_1 > 7$.

\begin{acknowledgements}
The work of YAF has been done in part under the auspices of the
Aix-Marseille~I University in 1998 and 1999.
YAF acknowledges also the support from the Russian Foundation for Basic
Research (grant 98--02--16734).
\end{acknowledgements}


\end{document}